\documentclass[nojss]{jss}\usepackage[]{graphicx}\usepackage[]{color}
\makeatletter
\def\maxwidth{ %
  \ifdim\Gin@nat@width>\linewidth
    \linewidth
  \else
    \Gin@nat@width
  \fi
}
\makeatother

\definecolor{fgcolor}{rgb}{0.345, 0.345, 0.345}
\newcommand{\hlnum}[1]{\textcolor[rgb]{0.686,0.059,0.569}{#1}}%
\newcommand{\hlstr}[1]{\textcolor[rgb]{0.192,0.494,0.8}{#1}}%
\newcommand{\hlcom}[1]{\textcolor[rgb]{0.678,0.584,0.686}{\textit{#1}}}%
\newcommand{\hlopt}[1]{\textcolor[rgb]{0,0,0}{#1}}%
\newcommand{\hlstd}[1]{\textcolor[rgb]{0.345,0.345,0.345}{#1}}%
\newcommand{\hlkwb}[1]{\textcolor[rgb]{0.69,0.353,0.396}{#1}}%
\newcommand{\hlkwc}[1]{\textcolor[rgb]{0.333,0.667,0.333}{#1}}%
\newcommand{\hlkwd}[1]{\textcolor[rgb]{0.737,0.353,0.396}{\textbf{#1}}}%

\usepackage{framed}
\makeatletter
\newenvironment{kframe}{%
 \def\at@end@of@kframe{}%
 \ifinner\ifhmode%
  \def\at@end@of@kframe{\end{minipage}}%
  \begin{minipage}{\columnwidth}%
 \fi\fi%
 \def\FrameCommand##1{\hskip\@totalleftmargin \hskip-\fboxsep
 \colorbox{shadecolor}{##1}\hskip-\fboxsep
     \hskip-\linewidth \hskip-\@totalleftmargin \hskip\columnwidth}%
 \MakeFramed {\advance\hsize-\width
   \@totalleftmargin\z@ \linewidth\hsize
   \@setminipage}}%
 {\par\unskip\endMakeFramed%
 \at@end@of@kframe}
\makeatother

\definecolor{shadecolor}{rgb}{.97, .97, .97}
\definecolor{messagecolor}{rgb}{0, 0, 0}
\definecolor{warningcolor}{rgb}{1, 0, 1}
\definecolor{errorcolor}{rgb}{1, 0, 0}
\newenvironment{knitrout}{}{} 

\usepackage{alltt}

\usepackage{thumbpdf,lmodern}

\usepackage{framed}

\usepackage{amsmath}
\usepackage{amssymb}
\usepackage{amsfonts}
\usepackage{subfigure}
\usepackage{booktabs}
\usepackage{caption}

\newcommand{\bA}{\textbf{A}}
\newcommand{\bb}{\textbf{b}}
\newcommand{\bB}{\textbf{B}}

\newcommand{\bC}{\textbf{C}}

\newcommand{\bD}{\textbf{D}}

\newcommand{\bK}{\textbf{K}}

\newcommand{\bI}{\textbf{I}}

\newcommand{\bG}{\textbf{G}}

\newcommand{\bL}{\textbf{L}}

\newcommand{\bO}{\textbf{O}}

\newcommand{\bs}{\textbf{s}}
\newcommand{\bS}{\textbf{S}}
\newcommand{\bt}{\textbf{t}}

\newcommand{\bu}{\textbf{u}}
\newcommand{\bU}{\textbf{U}}
\newcommand{\bv}{\textbf{v}}
\newcommand{\bV}{\textbf{V}}
\newcommand{\bw}{\textbf{w}}
\newcommand{\bW}{\textbf{W}}

\newcommand{\bX}{\textbf{X}}
\newcommand{\by}{\textbf{y}}

\newcommand{\bZ}{\textbf{Z}}
\newcommand{\bz}{\textbf{z}}

\newcommand{\mbs}[1]{\boldsymbol{#1}}

\newcommand{\bSigma}{\mbs{\Sigma}}

\newcommand{\bphi}{\mbs{\phi}}
\newcommand{\bmu}{\mbs{\mu}}
\newcommand{\bbeta}{\mbs{\beta}}

\newcommand{\btheta}{{\mbs{\theta}}}
\newcommand{\btau}{{\mbs{\tau}}}

\newcommand{\ben}{\begin{equation*}}
\newcommand{\een}{\end{equation*}}
\newcommand{\bean}{\begin{eqnarray*}}
\newcommand{\eean}{\end{eqnarray*}}
\newcommand{\bsm}{\begin{smallmatrix}}
\newcommand{\esm}{\end{smallmatrix}}
\newcommand{\bmat}{\begin{matrix}}
\newcommand{\emat}{\end{matrix}}

\newcommand{\tbeta}{\tilde{\beta}}

\newcommand{\bzero}{\textbf{0}}
\newcommand{\bones}{\textbf{1}}
\newcommand{\given}{\,|\,}

\newcommand{\chol}{\mbox{\texttt{chol}}}
\newcommand{\trsolve}{\mbox{\texttt{trsolve}}}

\author{Andrew O. Finley\\Michigan State University
   \And Sudipto Banerjee\\University of California, Los Angeles}
\Plainauthor{Andrew O. Finley, Sudipto Banerjee}

\title{Bayesian spatially varying coefficient models in the \pkg{spBayes} \proglang{R} package}
\Plaintitle{Bayesian spatially varying coefficient models in the spBayes R package}
\Shorttitle{Spatially varying coefficient in \proglang{R}}

\Abstract{
  This paper describes and illustrates new functionality for fitting spatially varying coefficients models in the \pkg{spBayes} (version 0.4-2) \proglang{R} package. The new \code{spSVC} function uses a computationally efficient Markov chain Monte Carlo algorithm and extends current \pkg{spBayes} functions, that fit only space-varying intercept regression models, to fit independent or multivariate Gaussian process random effects for any set of columns in the regression design matrix. Newly added \proglang{OpenMP} parallelization options for \code{spSVC} are discussed and illustrated, as well as helper functions for joint and point-wise prediction and model fit diagnostics. The utility of the proposed models is illustrated using a PM$_{10}$ analysis over central Europe.
}

\Plainkeywords{MCMC, multivariate Gaussian Process, kriging, R}

\Address{
  Andrew Finley\\
  Department of Forestry\\
  Michigan State University\\
  Natural Resources Building\\
  480 Wilson Road, Room 126\\
  East Lansing, MI 48824-6402\\
  E-mail: \email{finleya@msu.edu}\\
  URL: \url{https://www.finley-lab.com}\\
  \\
  Sudipto Banerjee\\
  Fielding School of Public Health\\
  University of California, Los Angeles\\
  650 Charles E. Young Dr. South\\
  Los Angeles, CA 90095-1772\\
  E-mail: \email{sudipto@ucla.edu}\\
  URL: \url{https://ph.ucla.edu/faculty/banerjee}
}
\IfFileExists{upquote.sty}{\usepackage{upquote}}{}
\begin{document}

\section{Introduction} \label{sec:intro}

In this paper we describe and illustrate extended functionality of a recent reformulation and rewrite of core functions in the \pkg{spBayes} \citep{FBG15} \proglang{R} \citep{R} package. The \pkg{spBayes} package provides a suite of univariate and multivariate regression models for both Gaussian and non-Gaussian outcomes that are spatially indexed. There are, by now, many \proglang{R} packages that provide similar functionality. A recent read of the ``Analysis of Spatial Data'' CRAN Task View \citep{CRANSP} yielded $\sim$46 packages listed for geostatistical analysis---and this is not an exhaustive accounting of packages available for such analyses. \cite{FBG15} focused on laying out computationally efficient and flexible MCMC algorithms for estimating an array of spatio-temporal Gaussian process (GP) models. However, while the proposed sampling algorithms were quite general, only a narrow set of models were implemented in \pkg{spBayes}. Specifically, users could only specify univariate or multivariate GPs on model intercepts. Now, the addition of the \code{spSVC} function to \pkg{spBayes} (version 0.4-2 available on CRAN 3/7/2019) aims to provide additional user options for placing univariate or multivariate GPs on any set of model regression coefficients. 

Such functionality is not unique, there are several \proglang{R} packages capable of fitting spatially varying coefficient (SVC) models. Some are specifically designed to work with spatial or spatio-tempral data and others provide flexibility to allow coefficients to vary by some generic set of variables, which could be indexes in a coordinate system. Most of these packages employ some flavor of spline or kernel based regression method to allow varying impact of predictors. \cite{Hastie93} and \cite{Fan08} offer a general development of varying coefficient models and \cite{Gelfand03} provide treatment particular to spatial settings. Regarding implementation in \proglang{R}, the \pkg{spgwr} package \citep{spgwr} implements geographically weighted regression as originally detailed in \cite{Fotheringham02}. Key spline-based package options include \pkg{mgcv} \citep{wood04}, \pkg{svcm} \citep{svcm}, \pkg{np} \citep{np}, and \pkg{mboost} \citep{mboost}. \cite{vcrpart} recently developed a tree-based varying coefficient model (TVCM) algorithm and associated \pkg{vcrpart} package. The packages \pkg{walker} \citep{walker, walker-arxiv}, \pkg{spTimer} \cite{spTimer}, and \pkg{spTDyn} \citep{spTDyn, spTDyn-JSCS, spTDyn-JRSS} offer Bayesian time and space-time SVC models. Other Bayesian options include model development using more general software such as INLA \citep{INLA, Lindgren15, Bakka19} and Stan \citep{stan, RStan}, which can be called from their respective \proglang{R} packages. 

The \code{spSVC} function offers Markov chain Monte Carlo (MCMC) based SVC inference using an efficient sampling algorithm. The algorithm's efficiency derives from updates to only covariance parameters (i.e., regression coefficients and random effects are integrated out), computing parallelization, and use of tuned and/or multi-threaded matrix algebra libraries. Subsequent sections define the model and algorithm specifics, software features, and illustrative analyses of simulated and real data.

\section{Models and software} \label{sec:models}
\noindent Let $y(\bs)$ be the dependent variable (response or outcome) at location $\bs$ and consider the spatially varying regression model,
\begin{equation}\label{eq: spatially_varying_regression}
 y(\bs) = (\beta_1 + \delta_1w_1(\bs)) + \sum_{j=2}^{p} x_j(\bs)\left\{\beta_j + \delta_jw_j(\bs)\right\} + \epsilon(\bs)\;,
\end{equation}
where $x_j(\bs)$, for each $j=2,\ldots,p$ with $p\geq 1$, is the known value of a predictor at location $\bs$, $\beta_j$ is the regression coefficient corresponding to $x_j(\bs)$, $\beta_1$ is an intercept, and $\epsilon(\bs)$ is a Gaussian measurement error process independently distributed for each $\bs$. The quantities $w_1(\bs)$ and $w_j(\bs)$ are spatial processes corresponding to the intercept and predictors, thereby yielding a spatially varying regression model. We further accommodate the possibility that not all the predictors will have spatially varying impact on the outcome. Thus, $\delta$'s in (\ref{eq: spatially_varying_regression}) are binary indicators assuming the value $1$ if the associated predictor has a spatially varying regression coefficient and $0$ otherwise. For later convenience, when the respective $\delta = 1$ we define $\tbeta_1(\bs) = \beta_1 + \delta_1w_1(\bs)$ and $\tbeta_j(\bs) =  \beta_j + \delta_jw_j(\bs)$ as the space-varying regression coefficients.

Let ${\cal S} = \{\bs_1,\bs_2,\ldots,\bs_n\}$ be the set of spatial locations from which $y(\bs_i)$ and the predictors have been observed. Let $\bw$ be the $nr\times 1$ vector obtained by stacking up $\bw(\bs_i)$'s, where each $\bw(\bs_i)$ is an $r\times 1$ vector with $j$-th entry $w_j(\bs_i)$, $j=1,2,\ldots,r$ and $i=1,2,\ldots,n$. We treat $\bw(\bs)$ as a multivariate Gaussian process \citep[see, e.g.,][]{BCG14} so the matrix $\bK_{\theta}$ is an $nr\times nr$  spatial covariance matrix constructed as a block matrix with $(i,j)$-th block obtained from the $r\times r$ cross-covariance matrix $\bK_{\theta}(\bs_i,\bs_j)$ specifying the multivariate spatial process $\bw(\bs)$. In addition, $\bbeta$ is the $p\times 1$ regression coefficient corresponding to $\bX$, and $\btheta$ and $\btau$ are the parameters in $\bK_{\theta}$ and $\bD_{\tau}$, respectively.

Consider the Bayesian hierarchical model built from (\ref{eq: spatially_varying_regression}),
\begin{align}\label{eq: spatially_varying_regression_bhm}
 p(\btau,\btheta) \times N(\bbeta \given \bmu_{\beta}, \bSigma_{\beta})\times N(\bw\given \bzero, \bK_{\theta}) \times N(\by \given \bX\bbeta + \bZ\bw, \bD_{\tau})\;, 
\end{align}
where $\by$ is $n\times 1$ with $i$-th element $y(\bs_i)$, $\bX$ is $n\times p$ with the first column $\bones$ and the remaining $p-1$ columns corresponding to the predictors $x_j(\bs_i)$ in (\ref{eq: spatially_varying_regression}). The matrix $\bZ$ is $n\times nr$, where $r\leq p$, with precisely those columns of $\bX$ which have $\delta_j=1$. 

We will offer users the option to scale and center the matrix $\bX$. Note that $\bZ$ is constructed from $\bX$, and thus, for a scaled and centered $\bX$, the predictors used in $\bZ$ will also be scaled and centered. Scaling and centering often improves numerical stability and provides more robust estimation of spatially varying regression models \citep[see, e.g.,][]{GKSB03}.

Some further specifications are in order. In \code{spSVC} we will fix $\bD_{\tau} = \tau^2\bI_n$ so $\btau = \{\tau^2\}$ is the scalar quantity representing the measurement error variance or ``nugget'' in geostatistics. The cross-covariance matrix $\bK_{\theta}(\bs,\bt)$, where $\bt$ is a generic location, will most generally be modeled using the Linear Model of Coregionalization (LMC). Here, we will model $\bK_{\theta}(\bs,\bt) = \bA\Gamma(\bs,\bt)\bA^{\top}$, where $\bA$ is an $r\times r$ lower triangular matrix and $\Gamma(\bs,\bt)$ is a diagonal matrix with $\rho_{j}(\bs,\bt)$ being the $j$-th diagonal element, where $\rho_{j}(\bs,\bt)$ is a spatial correlation function with parameters specific to $w_j(\bs)$. Here, $\btheta$ in (\ref{eq: spatially_varying_regression_bhm}) corresponds to $\{\bA, \{\bphi_j\}_{j=1}^{r}\}$ where each $\bphi_j$ is a collection of parameters in the spatial correlation function. For example, with the Mat\'{e}rn covariance function each $\bphi_j$ comprises a spatial decay parameter and a smoothness parameter. 

The covariance structure for $\bw(\bs)$ within any location $\bs$ is captured by $\bA\bA^{\top}$, which identifies with the Cholesky decomposition for $\mbox{var}\{\bw(\bs)\}$. In general, we will specify priors as
\begin{equation}\label{eq: priors_process_nugget}
 p(\tau^2,\btheta) = IG(\tau^2\given a_{\tau}, b_{\tau}) \times IW(\bA\bA^{\top}\given r_{a}, \bS_{a}) \times \prod_{j=1}^r p(\bphi_j)\;,
\end{equation}
where $IG$ is inverse-Gamma, $IW$ is inverse-Wishart, and each $p(\bphi_j)$ can be one of the several distributions provided by \pkg{spBayes}. Another particular choice offered by \code{spSVC} specifies $\bA = \mbox{diag}(\sigma_1,\ldots,\sigma_r)$, so that $\bK_{\theta}(\bs,\bt)$ is diagonal with entries $\sigma_j^2\rho_j(\bs,\bt)$, in which case we assume $IG(\sigma_j^2\given a_{\sigma}, b_{\sigma})$ for $j=1,2,\ldots,r$. Choices for $\rho_j$ include any of the standard correlation functions offered by \pkg{spBayes}.  


\subsection{Parameter estimation and computational considerations}

Bayesian inference for (\ref{eq: spatially_varying_regression}) involves sampling the parameters $\btheta$, $\bbeta$ and $\bw$ from their marginal posterior distributions. Such sampling algorithms require expensive operations on dense matrices such as decomposition and multiplication. Therefore, as we have outlined below, care is needed to use efficient numerical algorithms such as Cholesky factorizations, working with triangular systems, and avoiding redundant operations.  

\subsubsection{Sampling the process parameters}\label{Sec: Sampling_Process_Parameters_Generic}
Sampling from (\ref{eq: spatially_varying_regression_bhm}) employs MCMC methods, in particular Gibbs sampling and random walk Metropolis steps (e.g., \citealt{robert2010}). For faster convergence, we integrate out $\bbeta$ and $\bw$ from the model and first sample from $\displaystyle p(\btheta\given \by) \propto p(\btheta) \times N(\by\given \bX\bmu_{\beta}, \bSigma_{y\given\theta})$, where $\bSigma_{y\given\theta} = \bX\bSigma_{\beta}\bX^\top + \bZ\bK_{\theta}\bZ^\top + \bD_{\tau}$. This matrix needs to be constructed for every update of $\btheta$. $\bD_{\tau}$ is diagonal and $\bX\bSigma_{\beta}\bX^\top$ is fixed, so the computation involves the matrix $\bZ\bK_{\theta}\bZ^\top$ which requires $rn^2$ flops (floating point operations).  

We adopt a random-walk Metropolis step with a multivariate normal proposal (same dimension as there are parameters in $\btheta$) after transforming parameters to have support over the entire real line. This involves evaluating
\begin{equation}\label{Eq:target_theta_generic}
\log p(\btheta\given\by) = \mbox{const} + \log p(\btheta) - \frac{1}{2} \log |\bSigma_{y\given\theta}| - \frac{1}{2}Q(\btheta)\; ,
\end{equation}
where $Q(\btheta) = (\by - \bX\bmu_{\beta})^\top\bSigma^{-1}_{y\given \theta}(\by-\bX\bmu_{\beta})$. Generally, we compute $\bL=\mbox{\texttt{chol}}(\bSigma_{y\given \theta})$, where $\chol(\bSigma_{y\given \theta})$ returns the lower-triangular Cholesky factor $\bL$ of $\bSigma_{y\given\btheta}$. This involves $O(n^3/3)$ flops. Next, we obtain $\bu = \trsolve(\bL, \by - \bX\bmu_{\beta})$, which solves the triangular system $\bL\bu = \by - \bX\bmu_{\beta}$. This involves $O(n^2)$ flops and $Q(\btheta) = \bu^\top\bu$ requires another $2n$ flops. The log-determinant in (\ref{Eq:target_theta_generic}) is evaluated as $2\sum_{i=1}^n \log l_{i,i}$, where $l_{i,i}$ are the diagonal entries in $\bL$. Since $\bL$ has already been obtained, the log-determinant requires another $n$ steps. Therefore, the Cholesky factorization dominates the work and computing (\ref{Eq:target_theta_generic}) is achieved in $O(n^3)$ flops.

If $\bbeta$ is flat, i.e., $\bSigma_{\beta}^{-1}=\bO$, the analogue of distribution (\ref{Eq:target_theta_generic}) is 
\begin{equation}\label{Eq:target_theta_generic_beta_flat}
\log p(\btheta\given\by) = \mbox{constant} + \log p(\btheta) - \frac{1}{2} \log |\bX^\top\bSigma_{y\given\beta,\theta}^{-1}\bX| - \frac{1}{2} \log |\bSigma_{y\given\beta,\theta}| - \frac{1}{2}Q(\btheta),
\end{equation}
where $\bSigma_{y\given\beta,\theta} = \bZ\bK_{\theta}\bZ^\top + \bD_{\tau}$ and $Q(\btheta) = \by^\top\bSigma_{y\given\beta,\theta}^{-1}\by - \bb^\top(\bX^\top\bSigma_{y\given\beta,\theta}^{-1}\bX)^{-1}\bb$ and $\bb = \bX^\top\bSigma_{y\given\beta,\theta}^{-1}\by$. Computations proceed similar to the above. We first evaluate  $\bL = \mbox{\texttt{chol}}(\bSigma_{y\given \beta,\theta})$ and then obtain $[\bv:\bU] = \trsolve(\bL, [\by:\bX])$, so $\bL\bv=\by$ and $\bL\bU = \bX$. Next, we evaluate $\bW = \mbox{\texttt{chol}}(\bU^\top\bU)$, $\bb = \bU^\top\bv$ and solve $\tilde{\bb}=\trsolve(\bW,\bb)$. Finally, (\ref{Eq:target_theta_generic_beta_flat}) is evaluated as
 \begin{equation*}
    \log p(\btheta) - \sum^p_{i=1}\log w_{i,i} - \sum^n_{i=1} \log l_{i,i} - \frac{1}{2}(\bv^\top\bv - \tilde{\bb}^\top\tilde{\bb}),
  \end{equation*}
where $w_{i,i}$'s and $l_{i,i}$'s are the diagonal elements in $\bW$ and $\bL$ respectively. The number of flops is again of cubic order in $n$.

Importantly, our strategy above avoids computing inverses. We use Cholesky factorizations and solve only triangular systems. If $n$ is not large, say $\sim$$10^2$, this strategy is feasible. As described in Section~\ref{sec:features} and illustrated in Section~\ref{sec:illustrations}, use of efficient and parallelized numerical linear algebra routines yields substantial gains in computing time.
   
\subsubsection{Sampling the slope and the random effects}\label{Sec: Sampling_Slope_AND_Random_Effects}
Once we have obtained marginal posterior samples $\btheta$ from $p(\btheta\given \by)$, we can draw posterior samples of $\bbeta$ and $\bw$ using \emph{composition sampling}. Suppose $\{\btheta^{(1)},\btheta^{(2)},\ldots,\btheta^{(M)}\}$ are $M$ samples from $p(\btheta\given\by)$. Drawing $\bbeta^{(k)} \sim p(\bbeta\given \btheta^{(k)}, \by)$ and $\bw^{(k)}\sim p(\bw\given \btheta^{(k)},\by)$ for $k=1,2,\ldots M$ results in $M$ samples from $p(\bbeta\given\by)$ and $p(\bw\given\by)$ respectively. Only the samples of $\btheta$ obtained after convergence (i.e., \emph{post burn-in}) of the MCMC algorithm need to be stored. 

To elucidate further, note that $\bbeta\given \btheta,\by \sim N_p(\bB\bb, \bB)$ with mean $\bB\bb$ and variance-covariance matrix $\bB$, where
\begin{align}\label{Eq: Full_Conditional_beta_generic}
\bb = \bSigma_{\beta}^{-1}\bmu_{\beta} + \bX^\top\bSigma_{y\given \beta,\theta}^{-1}\by\; \mbox{ and }\; \bB = \left(\bSigma_{\beta}^{-1} + \bX^\top\bSigma_{y\given \beta, \theta}^{-1}\bX\right)^{-1}\;.
\end{align}
For each $k=1,2,\ldots,M$, we compute $\bB$ and $\bb$ at the current value $\btheta^{(k)}$ and draw $\bbeta^{(k)}\sim N_p(\bB\bb, \bB)$. This is achieved by computing $\bb=\bSigma_{\beta}^{-1}\bmu_{\beta} + \bU^\top\bv$, where $\bL = \chol(\bSigma_{y\given\beta,\theta^{(k)}})$ and $[\bv : \bU] =\trsolve(\bL, [\by : \bX])$. Next, we generate $p$ independent standard normal variables, collect them into $\bz$ and set
\begin{align}\label{Eq: Recover_beta_Generic}
 \bbeta^{(k)} = \trsolve\left(\bL_{B}^\top,\trsolve(\bL_{B}, \bb)\right) + \trsolve(\bL_{B}^\top,\bz)\; ,
\end{align}
where $\displaystyle \bL_{B} = \chol\left(\bSigma_{\beta}^{-1} + \bU^\top\bU\right)$. This completes the $k$-th iteration. After $M$ iterations, we obtain $\{\bbeta^{(1)}, \bbeta^{(2)},\ldots, \bbeta^{(M)}\}$, which are samples from $p(\bbeta\given\by)$. 

Mapping point or interval estimates of spatial random effects is often helpful in identifying missing regressors and/or building a better understanding of model adequacy. $\bSigma_{y\given w,\theta} = \bX\bSigma_{\beta}\bX^\top + \bD_{\tau}$ and note that  $\bw\given \btheta,\by \sim N(\bB\bb, \bB)$, where
\begin{align}\label{Eq: Full_Conditional_alpha_generic}
\bb = \bZ^\top\bSigma_{y\given w, \theta}^{-1}(\by-\bX\bmu_{\beta})\; \mbox{ and }\; \bB = \left(\bK_{\theta}^{-1} + \bZ^\top\bSigma_{y\given w, \theta}^{-1}\bZ\right)^{-1}\; .
\end{align}
The vector $\bb$ here is computed analogously as for $\bbeta$.  For each $k=1,2,\ldots,M$ we now evaluate $\bL = \chol(\bSigma_{y\given w,\theta^{(k)}})$, $[\bv : \bU] = \trsolve(\bL, [\by-\bX\bmu_{\beta}: \bZ(\btheta^{(k)})])$ and set $\bb = \bU(\btheta^{(k)})^\top\bv$. For computing $\bB$, one could proceed as for $\bbeta$ but that would involve $\chol(\bK(\btheta))$, which may become numerically unstable for certain covariance functions (e.g., the Gaussian or the Mat\'ern with large $\nu$). For robust software performance we define $\bG_{\theta}^{-1} = \bZ'\bSigma_{y\given w, \theta}^{-1}\bZ$ and utilize the identity \citep{henderson1981}
\[
 \left(\bK_{\theta}^{-1} + \bG_{\theta}^{-1}\right)^{-1} =  \bG_{\theta} - \bG_{\theta}\left(\bK_{\theta} + \bG_{\theta}\right)^{-1}\bG_{\theta}\;  
\]
to devise a numerically stable algorithm. For each $k=1,2,\ldots,M$, we evaluate $\bL = \chol(\bK_{\theta}^{(k)} + \bG_{\theta}^{(k)})$, $\bW = \trsolve(\bL, \bG_{\theta}^{(k)})$ and $\bL_{B} = \chol(\bG_{\theta}^{(k)} - \bW^\top\bW)$. If $\bz$ is a $r\times 1$ vector of independent standard normal variables, then we set
$\displaystyle \bw^{(k)} = \bL_{B}\bL_{B}^\top\bb + \bL_{B}\bz$. The resulting $\{\bw^{(1)}, \bw^{(2)},\ldots, \bw^{(M)}\}$ are samples from $p(\bw\given \by)$.
 
\subsubsection{Spatial predictions}\label{Sec: Spatial_Predictions_Kriging}  
To predict a random $n_0\times 1$ vector $\by_0$ associated with a $n_0\times p$ matrix of predictors, $\bX_0$, we assume that
\begin{align}\label{Eq: Prediction_Joint_Distribution_generic}
 \begin{bmatrix}\by\\ \by_0\end{bmatrix} \bigg| \bbeta, \btheta \sim N_{n_0+n}\left(\begin{bmatrix}\bX\\ \bX_0\end{bmatrix}\bbeta,\; \begin{bmatrix} \bC_{11}(\btheta) & \bC_{12}(\btheta) \\ \bC_{12}(\btheta)^\top & \bC_{22}(\btheta) \end{bmatrix}\right)\; ,
\end{align}
where $\bC_{11}(\btheta) = \bSigma_{y\given\beta, \theta}$, $\bC_{12}(\btheta)$ is the $n\times n_0$ cross-covariance matrix between $\by$ and $\by_0$, and $\bC_{22}(\btheta)$ is the variance-covariance matrix for $\by_0$. A valid \emph{joint} distribution will supply a conditional distribution $p(\by_0\given\by,\bbeta,\btheta)$, which is normal with mean and variance
\begin{align}\label{Eq: Prediction_Conditional_Distribution_generic}
 \bmu_{p} &= \bX_0\bbeta + \bC_{12}(\btheta)^\top\bC_{11}(\btheta)^{-1}(\by-\bX\bbeta)\mbox{ and } \bSigma_{p} = \bC_{22}(\btheta) - \bC_{12}(\btheta)^\top\bC_{11}(\btheta)^{-1}\bC_{12}(\btheta)\; 
\end{align}
Bayesian prediction proceeds by sampling from the posterior predictive distribution $\displaystyle p(\by_0\given \by) = \int p(\by_0\given \by, \bbeta,\btheta)p(\bbeta,\btheta\given\by)d\bbeta d\btheta$. For each posterior sample of $\{\bbeta,\btheta\}$, we draw a corresponding $\by_0\sim N(\bmu_p,\bSigma_p)$. This produces samples from the posterior predictive distribution.

The posterior predictive computations involve only the retained MCMC samples after convergence. For any posterior sample $\{\bbeta^{(k)},\btheta^{(k)}\}$, we solve $[\bu : \bV] = \trsolve(\bL, [\by-\bX\bbeta^{(k)} : \bC_{12}(\btheta^{(k)})])$, where $\bL = \chol(\bC_{11}(\btheta^{(k)}))$. Next, we set $\bmu_p^{(k)} = \bX_0\bbeta^{(k)} + \bV^\top\bu$ and $\bSigma_p^{(k)} = \bC_{22}(\btheta^{(k)}) - \bV^\top\bV$ and draw $\by_0^{(k)}\sim N(\bmu_p^{(k)},\bSigma_p^{(k)})$.

Updating $\by_0^{(k)}$'s requires Cholesky factorization of $\bSigma_p$, which is $n_0\times n_0$ and can be expensive if $n_0$ is large. In most practical settings, it is sufficient to take $n_0=1$ and perform \emph{point-wise} predictions. 

\subsection{Software features}\label{sec:features}

The \code{spSVC} function accommodates the \code{spLM} function in \pkg{spBayes} and offers additional user options to simplify analysis and inference. The list below highlights some of these new options.

\begin{enumerate}
\item Any set of predictors can receive either independent univariate GPs or a multivariate GP.
  \item Prediction can be done by sampling from either the joint or point-wise (marginal) posterior predictive distribution.
\item \proglang{openMP} \citep{openmp98} support is available via the \code{n.omp.threads} argument for parameter estimation, composition sampling, model fit diagnostics, and prediction functions.
\item Matrix operation parallelization is available via multi-threaded implementations of Basic Linear Algebra Subprograms (BLAS; \url{www.netlib.org/blas}) and Linear Algebra Package (LAPACK; \url{www.netlib.org/lapack}).
  \item Coordinate system used to index observed and prediction locations can be of arbitrary dimension---users were previously restricted to using 2-dimensional systems.
    \item Univariate and multivariate random effect samples and space-varying coefficients are returned as lists with element names corresponding to the given predictor.
\end{enumerate}

\section{Illustrations}\label{sec:illustrations}

We consider two analyses to illustrate key features of \code{spSVC} along with supporting functions. The first analysis is of a simulated dataset and second is of an air pollution dataset that was previously analyzed in \cite{hamm2015} and \cite{datta2016}.

\subsection{Analysis of simulated data}
The simulated data \code{mvSVCData} is available in \pkg{spBayes} and comprises $n$=500 observations distributed within a 2-dimensional unit square spatial domain. At generic location $\bs$ the outcome was generated following 
\begin{equation}
  y(\bs) = \beta_0 + w_0(\bs) + a(\bs) \left\{\beta_a + w_a(\bs)\right\} + b(\bs)\left\{\beta_b + w_b(\bs)\right\} + \epsilon(\bs)\;.
\end{equation}

\begin{figure}[ht!]
\centering
\subfigure[True $w_0$]{\includegraphics[trim={0cm 0cm 0cm 2cm},clip,width=2in]{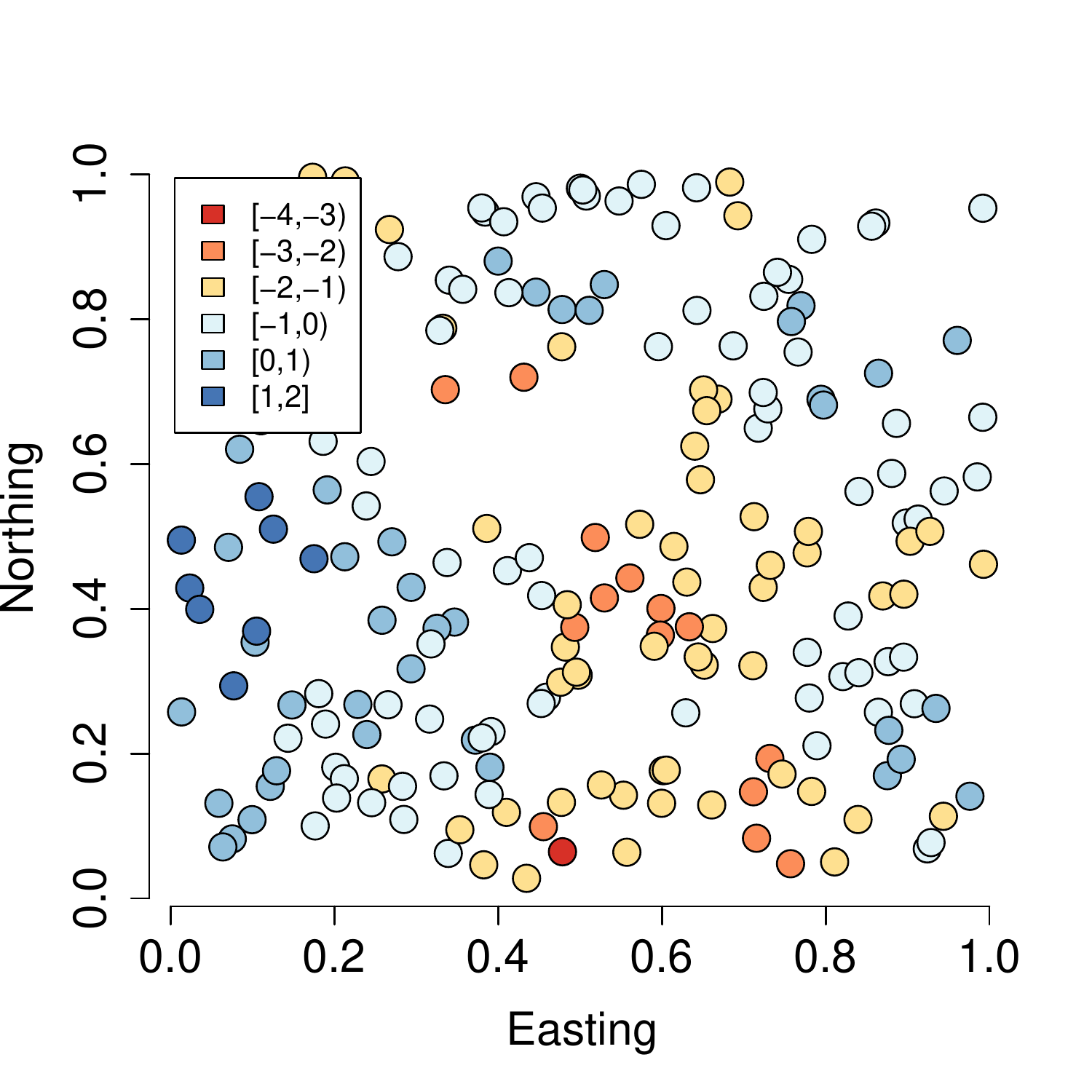}\label{w0True}} 
\subfigure[True $w_a$]{\includegraphics[trim={0cm 0cm 0cm 2cm},clip,width=2in]{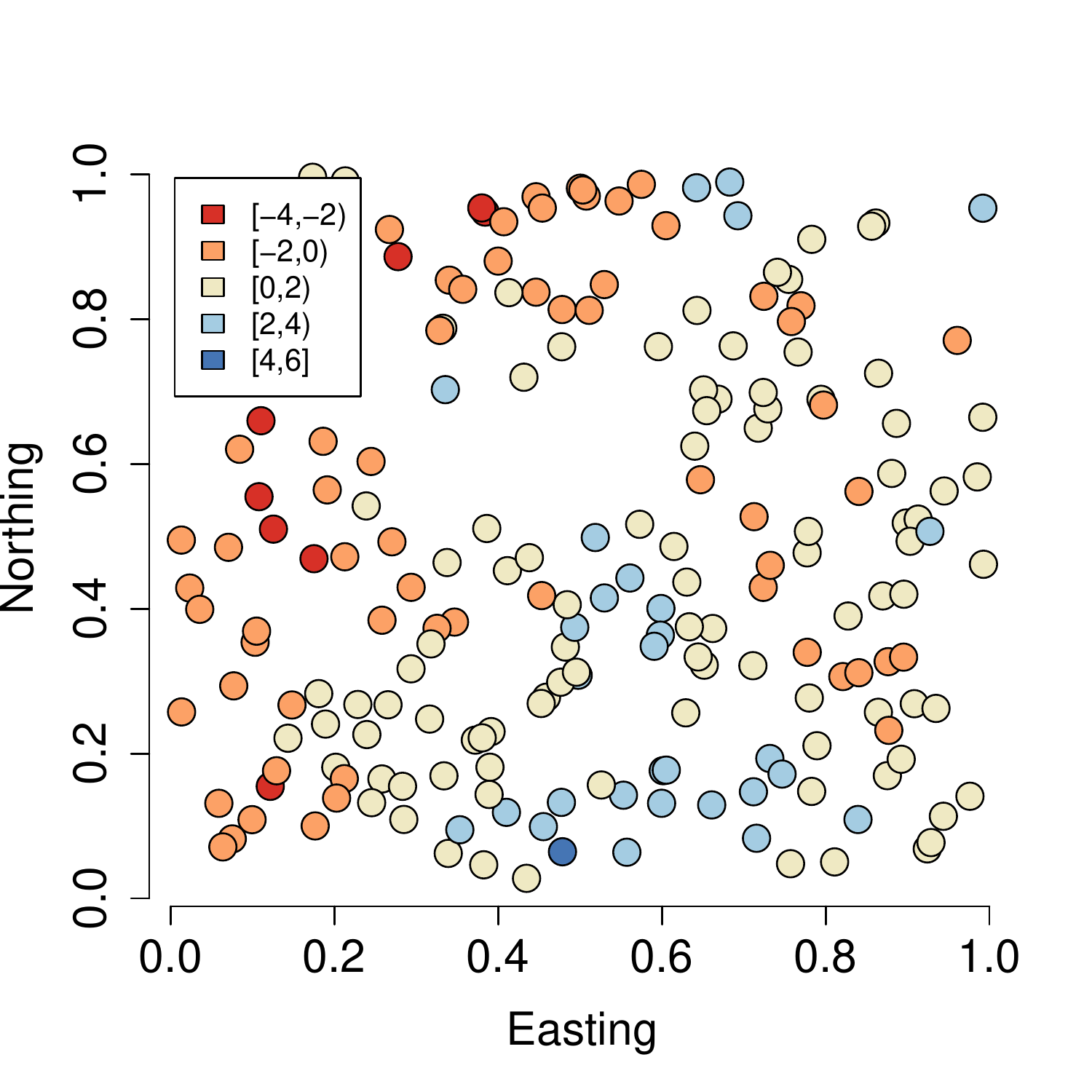}\label{waTrue}} 
\subfigure[True $w_b$]{\includegraphics[trim={0cm 0cm 0cm 2cm},clip,width=2in]{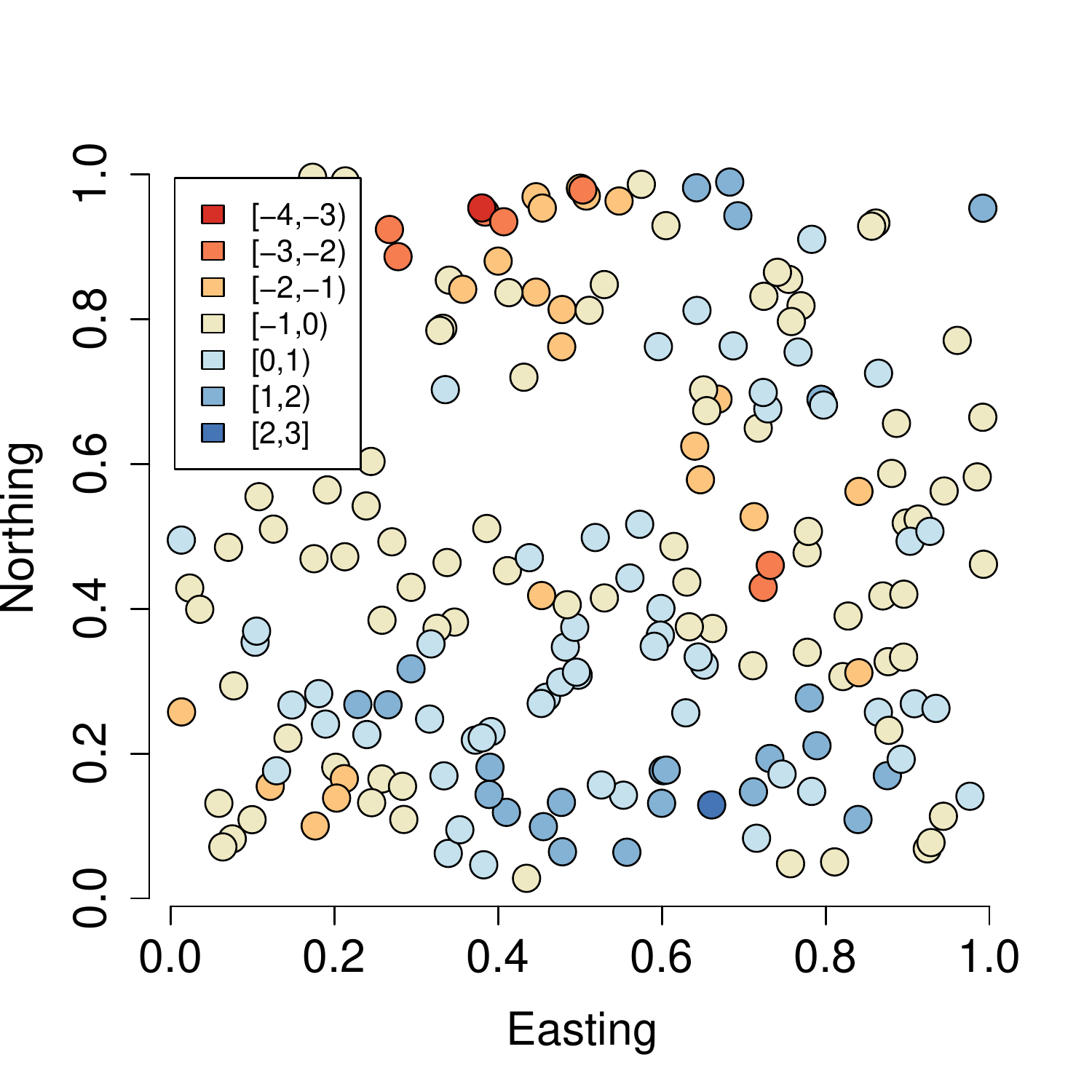}\label{wbTrue}}\\
\subfigure[True vs. fitted $w_0$]{\includegraphics[trim={0cm 0cm 0cm 2cm},clip,width=2in]{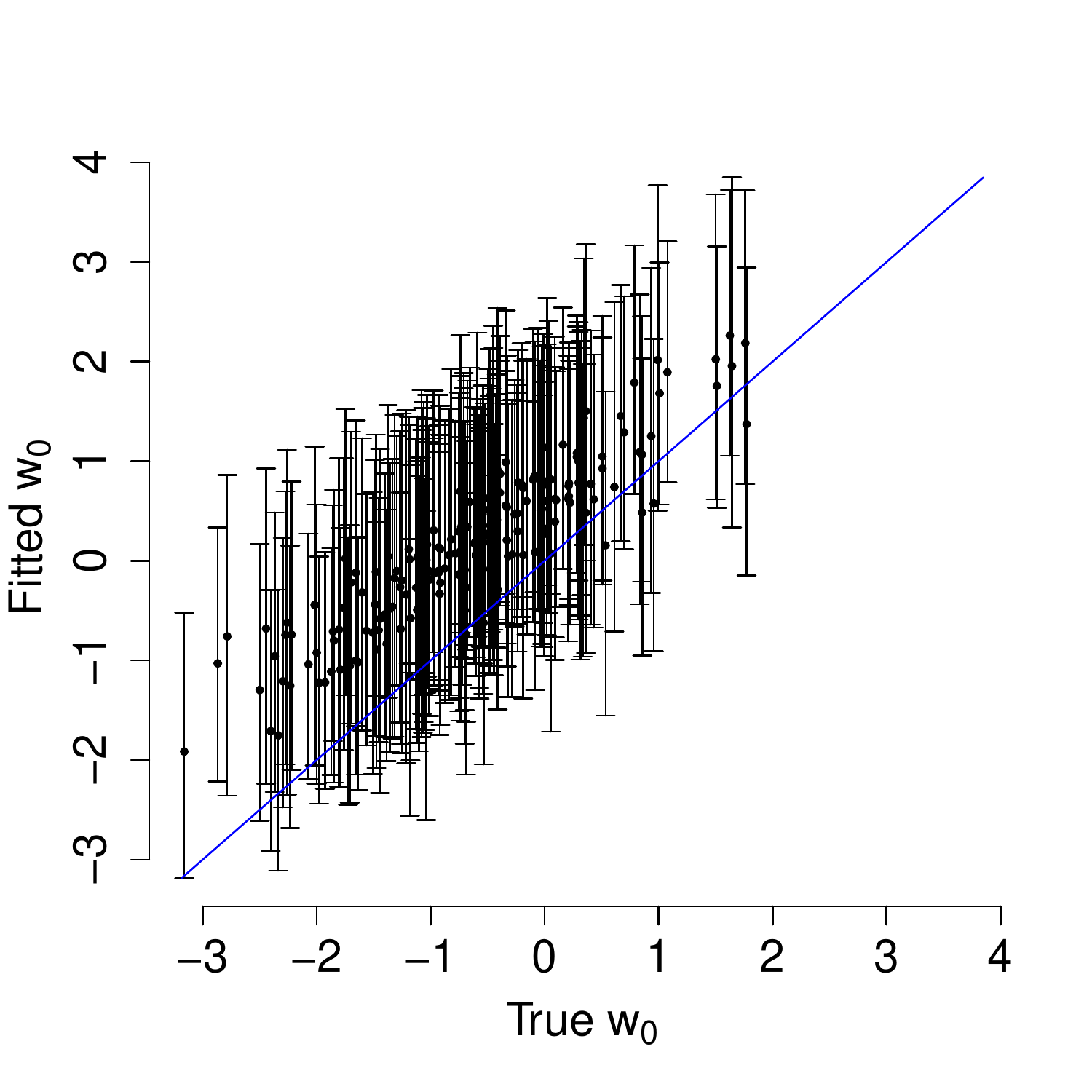}\label{w0TrueVsFitted}} 
\subfigure[True vs. fitted $w_a$]{\includegraphics[trim={0cm 0cm 0cm 2cm},clip,width=2in]{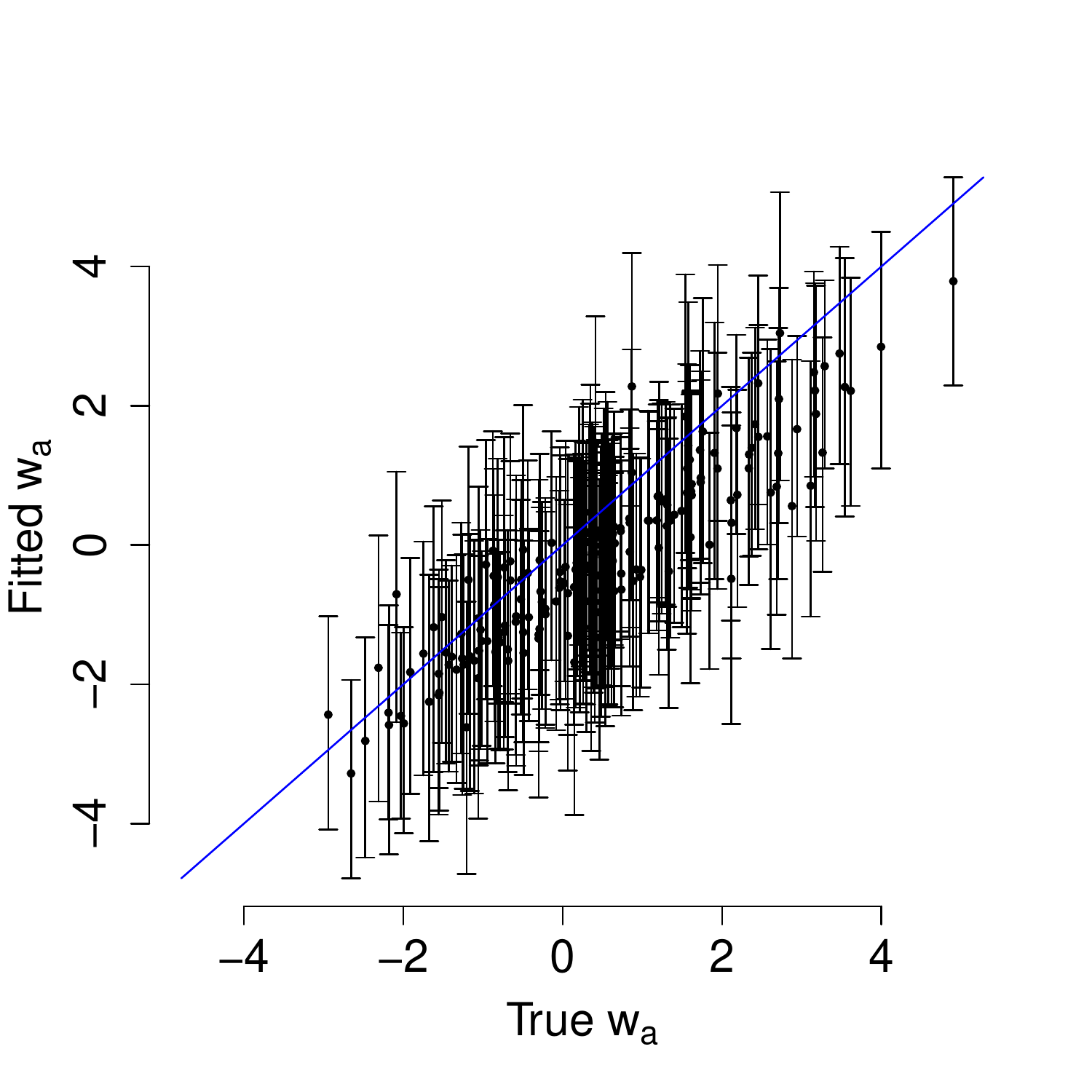}\label{waTrueVsFitted}} 
\subfigure[True vs. fitted $w_b$]{\includegraphics[trim={0cm 0cm 0cm 2cm},clip,width=2in]{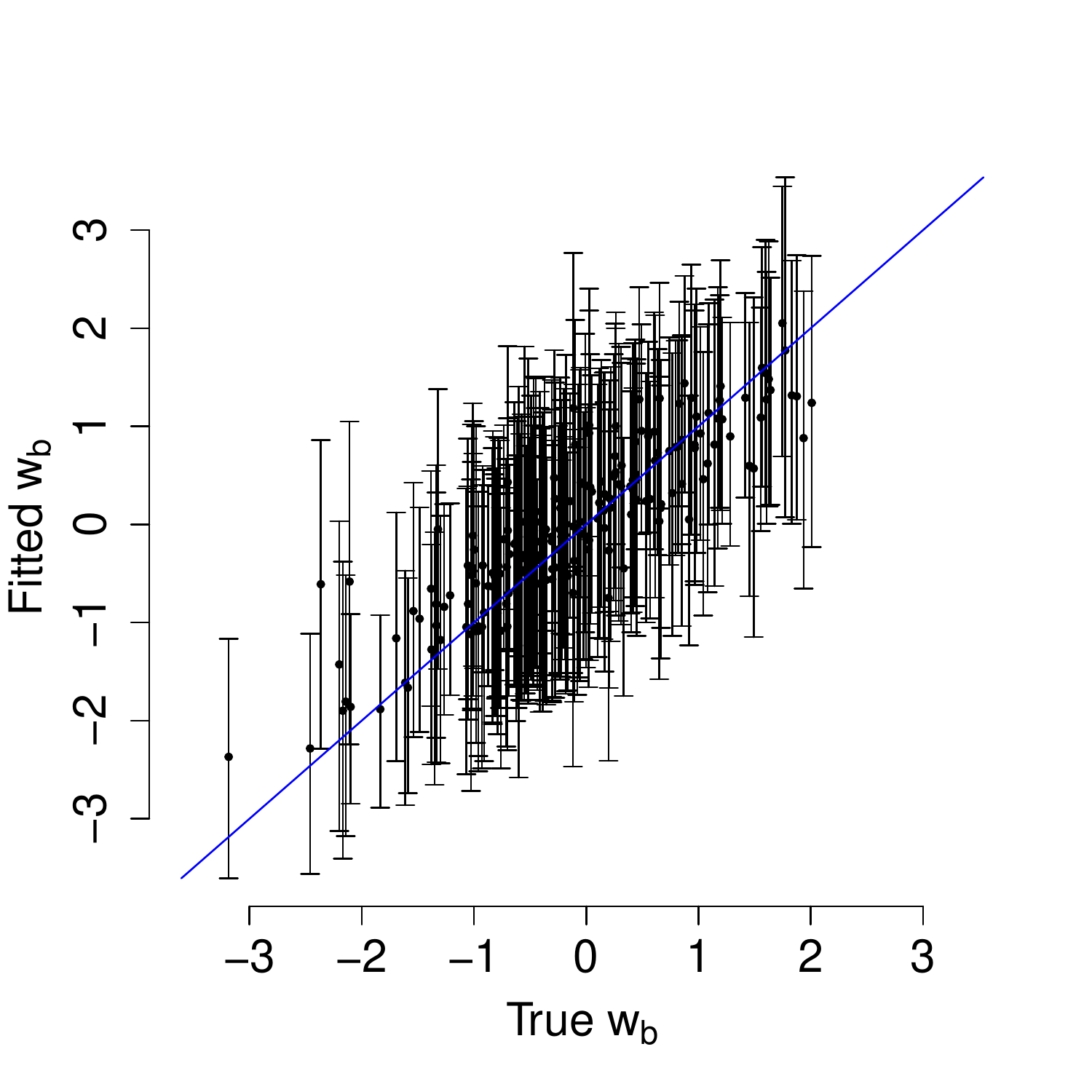}\label{wbTrueVsFitted}}
\caption{Simulated data analysis, observed (true) and model estimated (fitted) random effect values. Figures~\subref{w0TrueVsFitted}-\subref{wbTrueVsFitted} posterior median and 95\% credible interval shown as point and bars, respectively. }\label{simW}
\end{figure}

The predictors $a(\bs)$ and $b(\bs)$ were drawn from independent normal distributions with mean zero and variance one. The regression coefficients $\beta_0$, $\beta_a$, and $\beta_b$ equaled 1, 10, and -10, respectively. The $r$=3 spatial random effects associated with the intercept and predictors were generated from a non-separable multivariate GP. The cross-covariance function used to construct the $(i,j)$-th $r\times r$ block in the multivariate GP's $nr\times nr$ covariance matrix, i.e.,  $\bK_{\theta}(\bs_i,\bs_j) = \bA\Gamma(\bs_i,\bs_j)\bA^{\top}$, was 
\begin{equation}
  \begin{pmatrix}
    1  &0  &0\\
    -1  &1  &0\\
    0  &1  &0.1\\
  \end{pmatrix}
  \begin{pmatrix}
    \text{exp}(-4 d_{i,j})  &0  &0\\
    0&\text{exp}(-6 d_{i,j}) &0\\
    0&  0 &\text{exp}(-6 d_{i,j})\\
  \end{pmatrix}
  \begin{pmatrix}
    1  &-1  &0\\
    0  &1  &1\\
    0  &0  &0.1\\
  \end{pmatrix},
\end{equation}
where $d_{i,j}$ is the euclidean distance between location $\bs_i$ and $\bs_j$, and diagonal elements of $\Gamma(\bs_i,\bs_j)$ are the exponential correlation function $\text{exp}(-\phi_k d_{i,j})$ for $k=1,2,\ldots,r$. Figure~\ref{w0True}-\subref{wbTrue} display the realizations of $\bw_0$, $\bw_a$, and $\bw_b$. The residual term $\epsilon(\bs)$ was simulated from a Normal distribution with mean zero and variance $\tau^2 = 0.1$.

The code below specifies the model covariance parameters' prior distributions, and MCMC sampler starting and Metropolis proposal variance values. Here, we use a Uniform prior for the spatial decay parameters each with support from 1 to 10. The prior for the cross-covaraince matrix is an IW with degrees of freedom $r$ and identity scale matrix. The prior for the measurement error (or nugget variance) follows an IG with shape 2 and scale 1. 

The parameter priors, starting values, and Metropolis sampler proposal variances are passed to \code{spSVC} via the \code{priors}, \code{starting}, and \code{tuning} arguments, respectively. The proposed model is specified via the \code{formula} argument using syntax like that used in base \proglang{R}'s \code{lm}, with the addition of the \code{svc.cols} argument that accepts a vector of either integer indexes or character names to indicate the space-varying design matrix columns (i.e., the columns of $\bX$ with $\delta_j=1$). For example, in the call to \code{spSVC} below, the vector passed to \code{svc.cols} indicates we want the intercept and columns labeled \code{a} and \code{b} to follow a multivariate GP (or, equivalently, one could use the argument value \code{c(1,2,3)}).

\begin{knitrout}
\definecolor{shadecolor}{rgb}{0.969, 0.969, 0.969}\color{fgcolor}\begin{kframe}
\begin{alltt}
\hlkwd{data}\hlstd{(SVCMvData.dat)}

\hlstd{r} \hlkwb{<-} \hlnum{3}

\hlstd{n.ltr} \hlkwb{<-} \hlstd{r}\hlopt{*}\hlstd{(r}\hlopt{+}\hlnum{1}\hlstd{)}\hlopt{/}\hlnum{2}

\hlstd{priors} \hlkwb{<-} \hlkwd{list}\hlstd{(}\hlstr{"phi.Unif"}\hlstd{=}\hlkwd{list}\hlstd{(}\hlkwd{rep}\hlstd{(}\hlnum{1}\hlstd{,r),} \hlkwd{rep}\hlstd{(}\hlnum{10}\hlstd{,r)),}
               \hlstr{"K.IW"}\hlstd{=}\hlkwd{list}\hlstd{(r,} \hlkwd{diag}\hlstd{(}\hlkwd{rep}\hlstd{(}\hlnum{1}\hlstd{,r))),}
               \hlstr{"tau.sq.IG"}\hlstd{=}\hlkwd{c}\hlstd{(}\hlnum{2}\hlstd{,} \hlnum{1}\hlstd{))}

\hlstd{starting} \hlkwb{<-} \hlkwd{list}\hlstd{(}\hlstr{"phi"}\hlstd{=}\hlkwd{rep}\hlstd{(}\hlnum{3}\hlopt{/}\hlnum{0.5}\hlstd{,r),} \hlstr{"A"}\hlstd{=}\hlkwd{c}\hlstd{(}\hlnum{1}\hlstd{,}\hlnum{0}\hlstd{,}\hlnum{0}\hlstd{,}\hlnum{1}\hlstd{,}\hlnum{0}\hlstd{,}\hlnum{1}\hlstd{),} \hlstr{"tau.sq"}\hlstd{=}\hlnum{1}\hlstd{)}

\hlstd{tuning} \hlkwb{<-} \hlkwd{list}\hlstd{(}\hlstr{"phi"}\hlstd{=}\hlkwd{rep}\hlstd{(}\hlnum{0.1}\hlstd{,r),} \hlstr{"A"}\hlstd{=}\hlkwd{rep}\hlstd{(}\hlnum{0.01}\hlstd{, n.ltr),} \hlstr{"tau.sq"}\hlstd{=}\hlnum{0.01}\hlstd{)}

\hlstd{sim.m} \hlkwb{<-} \hlkwd{spSVC}\hlstd{(y}\hlopt{~}\hlstd{a}\hlopt{+}\hlstd{b,} \hlkwc{coords}\hlstd{=}\hlkwd{c}\hlstd{(}\hlstr{"x.coords"}\hlstd{,}\hlstr{"y.coords"}\hlstd{),} \hlkwc{data}\hlstd{=SVCMvData.dat,}
             \hlkwc{starting}\hlstd{=starting,} \hlkwc{svc.cols}\hlstd{=}\hlkwd{c}\hlstd{(}\hlstr{"(Intercept)"}\hlstd{,}\hlstr{"a"}\hlstd{,}\hlstr{"b"}\hlstd{),}
             \hlkwc{tuning}\hlstd{=tuning,} \hlkwc{priors}\hlstd{=priors,} \hlkwc{cov.model}\hlstd{=}\hlstr{"exponential"}\hlstd{,}
             \hlkwc{n.samples}\hlstd{=}\hlnum{10000}\hlstd{,} \hlkwc{n.report}\hlstd{=}\hlnum{5000}\hlstd{,} \hlkwc{n.omp.threads}\hlstd{=}\hlnum{4}\hlstd{)}
\end{alltt}
\begin{verbatim}
----------------------------------------
	General model description
----------------------------------------
Model fit with 200 observations.

Number of covariates 3.

Number of space varying covariates 3.

Using the exponential spatial correlation model.

Number of MCMC samples 10000.

Priors and hyperpriors:
	beta flat.
	K IW hyperpriors:
	df: 3.00000
	S:
	1.000	0.000	0.000	
	0.000	1.000	0.000	
	0.000	0.000	1.000	

	phi Unif lower bound hyperpriors:	1.000	1.000	1.000	
	phi Unif upper bound hyperpriors:	10.000	10.000	10.000	

	tau.sq IG hyperpriors shape=2.00000 and scale=1.00000

Source compiled with OpenMP, posterior sampling is using 4 thread(s).
-------------------------------------------------
		Sampling
-------------------------------------------------
Sampled: 5000 of 10000, 50.00%
Report interval Metrop. Acceptance rate: 34.64%
Overall Metrop. Acceptance rate: 34.64%
-------------------------------------------------
Sampled: 10000 of 10000, 100.00%
Report interval Metrop. Acceptance rate: 34.24%
Overall Metrop. Acceptance rate: 34.44%
-------------------------------------------------
\end{verbatim}
\end{kframe}
\end{knitrout}

As described in Section~\ref{sec:models}, \code{spSVC} computes and returns MCMC samples for only model covariance parameters. If \code{verbose=TRUE}, basic model specifications are written to the terminal followed by updates on the sampler's progress and Metropolis algorithm acceptance rate. The sampler progress report interval is controlled using the \code{n.report} argument. One should adjust the Metropolis sampler proposal variances to achieve an acceptance rate between $\sim$30-50\% \citep[see, e.g.,][for model fitting best practices]{gelman2013}. If it proves difficult to maintain an acceptable acceptance rate, the \code{amcmc} argument can be added to invoke an adaptive MCMC algorithm \citep{Roberts09} that automatically adjusts the tuning to achieve a target acceptance rate (see the manual page for more details).

The \code{n.omp.threads} argument in \code{spSVC} call above requests that key \code{for} loops within a given MCMC iteration use 4 threads via  \proglang{openMP} \citep{openmp98}. If the user's \proglang{R} is set up to use a parallelized version of BLAS then \code{n.omp.threads} will also control the number of threads in some LAPACK matrix operations. Such parallelization can greatly reduce the sampler's runtime. 

\begin{figure}[ht!]
\begin{center}
\includegraphics[trim={0cm 0cm 0cm 2cm},clip,width=4in]{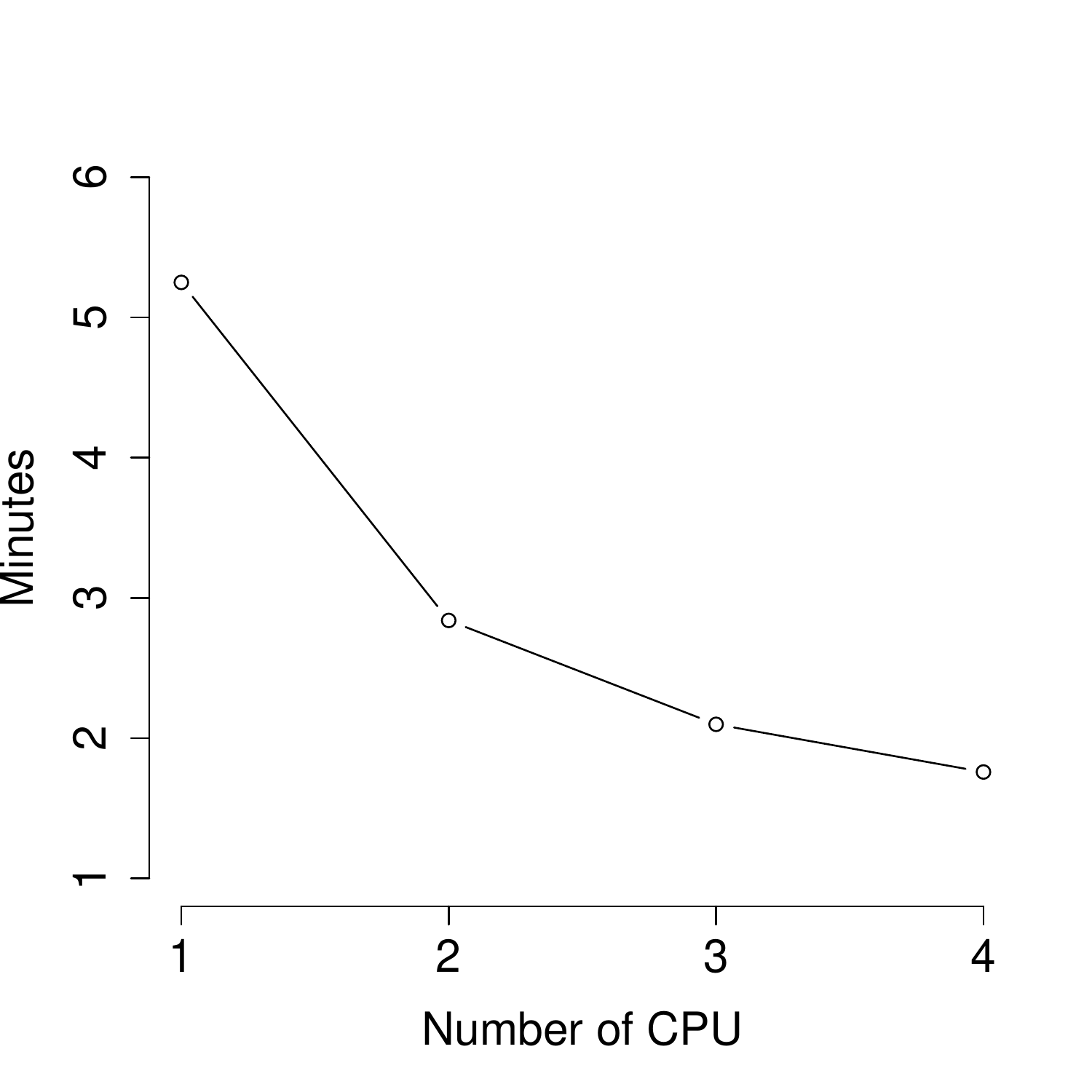}
\end{center}
\caption{Simulated data analysis assessment of optimal runtime.}\label{simRuntime}
\end{figure}

The computer used to conduct this analysis has an Intel(R) Core(TM) i7-8550U CPU @ 1.80GHz chip  with 4 cores and \proglang{R} compiled with \proglang{openMP}, as confirmed in the ``General model description'' printed after calling \code{spSVC}, which notes \texttt{Source compiled with OpenMP, posterior sampling is using 4 thread(s)}. \code{spSVC} will throw a warning if \proglang{R} was not compiled with \proglang{openMP} support and \code{n.omp.threads} is set to a value greater than 1. In addition to \proglang{openMP} support, the current implementation of \proglang{R} uses \texttt{openBLAS} \citep{zhang13} which is a version of BLAS capable of exploiting multiple processors. Figure~\ref{simRuntime} shows the runtime needed to complete 10000 MCMC iterations across the number of available CPUs.
  
Following execution of \code{spSVC}, the \code{sim.m} object holds MCMC samples for covariance parameters (\code{p.theta.samples}) along with data and model fitting details. Using, possibly post burn-in and thinned, \code{p.theta.samples}, the \code{spRecover} function conducts composition sampling to generate samples from the regression coefficients $\bbeta$ (\code{p.beta.recover.samples}), spatial random effects $\bw$ (\code{p.w.recover.samples}), and model fitted values (\code{p.y.samples}). \code{spRecover} also returns the subset of \code{p.theta.samples} (\code{p.theta.recover.samples}) used in the composition sampling. Further, for convenience, \code{spRecover} returns samples of the space-varying regression coefficients $\tbeta(\bs)$'s. \code{spRecover} appends these various composition sampling outputs to the \code{spSVC} input object, i.e., the \code{sim.m} object returned by \code{spRecover} below is identical to the \code{sim.m} object returned by \code{spSVC} except for the addition of the composition sampling results. In addition to providing posterior samples for all model parameters, a call to \code{spRecover} is necessary for subsequent prediction and model fit diagnostics, via \code{spPredict} and \code{spDiag} respectively. Like \code{spSVC}, \code{spRecover} takes advantage of multiple CPUs via \proglang{openMP} when available. 

\begin{knitrout}
\definecolor{shadecolor}{rgb}{0.969, 0.969, 0.969}\color{fgcolor}\begin{kframe}
\begin{alltt}
\hlstd{sim.m} \hlkwb{<-} \hlkwd{spRecover}\hlstd{(sim.m,} \hlkwc{start}\hlstd{=}\hlnum{5000}\hlstd{,} \hlkwc{thin}\hlstd{=}\hlnum{2}\hlstd{,} \hlkwc{n.omp.threads}\hlstd{=}\hlnum{4}\hlstd{,} \hlkwc{verbose}\hlstd{=}\hlnum{FALSE}\hlstd{)}
\end{alltt}
\end{kframe}
\end{knitrout}

\code{spSVC} and \code{spRecover} return samples as \code{coda} objects to simplify posterior summaries. Output below provides the post burn-in and thinned median with lower and upper 95\% credible bounds for $\beta$'s, cross-covariance matrix used to construct $\bK_{\theta}$, spatial decays $\phi$'s, and $\tau^2$. These summaries show the model captures well the parameter values used to simulate the data.  

\begin{knitrout}
\definecolor{shadecolor}{rgb}{0.969, 0.969, 0.969}\color{fgcolor}\begin{kframe}
\begin{alltt}
\hlkwd{round}\hlstd{(}\hlkwd{summary}\hlstd{(sim.m}\hlopt{$}\hlstd{p.beta.recover.samples)}\hlopt{$}\hlstd{quantiles[,}\hlkwd{c}\hlstd{(}\hlnum{3}\hlstd{,}\hlnum{1}\hlstd{,}\hlnum{5}\hlstd{)],}\hlnum{2}\hlstd{)}
\end{alltt}
\begin{verbatim}
               50%   2.5% 97.5%
(Intercept)   0.20  -1.05  1.16
a            10.53   9.37 11.92
b           -10.20 -11.08 -9.31
\end{verbatim}
\end{kframe}
\end{knitrout}

Note, following the notation in Section~\ref{sec:models} the cross-covaraince matrix used to simulated the data is
\begin{equation}
  \bK_{\theta}=\bA\bA^{\top}
  \begin{pmatrix}
    1  &0  &0\\
    -1  &1  &0\\
    0  &1  &0.1\\
  \end{pmatrix}
  \begin{pmatrix}
    1  &-1  &0\\
    0  &1  &1\\
    0  &0  &0.1\\
  \end{pmatrix}
  =  
  \begin{pmatrix}
    1  &-1  &0\\
    -1  &2  &1\\
    0  &1  &1.01\\
  \end{pmatrix}. 
\end{equation}

The posterior summary of the covariance parameters is below. Observed versus estimated random effects are given Figure~\ref{w0TrueVsFitted}-\subref{wbTrueVsFitted}.

\begin{knitrout}
\definecolor{shadecolor}{rgb}{0.969, 0.969, 0.969}\color{fgcolor}\begin{kframe}
\begin{alltt}
\hlkwd{round}\hlstd{(}\hlkwd{summary}\hlstd{(sim.m}\hlopt{$}\hlstd{p.theta.recover.samples)}\hlopt{$}\hlstd{quantiles[,}\hlkwd{c}\hlstd{(}\hlnum{3}\hlstd{,}\hlnum{1}\hlstd{,}\hlnum{5}\hlstd{)],}\hlnum{2}\hlstd{)}
\end{alltt}
\begin{verbatim}
                  50%  2.5% 97.5%
K[1,1]           1.11  0.58  2.78
K[2,1]          -1.00 -2.49 -0.46
K[3,1]          -0.06 -0.52  0.35
K[2,2]           1.89  1.17  3.60
K[3,2]           0.90  0.35  1.76
K[3,3]           1.14  0.63  2.03
tau.sq           0.18  0.10  0.33
phi.(Intercept)  4.16  1.57  8.65
phi.a            4.98  2.09  9.81
phi.b            6.13  1.55  9.89
\end{verbatim}
\end{kframe}
\end{knitrout}

\subsection{Analysis of air pollution data}\label{sec:dpm10}

Increases in human morbidity and mortality is a known outcome to airborne particulate matter (PM) exposure \citep{BrunekreefH02a,LoomisEtAl13a,HoekEtAl13a}. In response, regulatory agencies have instigated monitor programs and regulate PM concentrations. One such regulation by the European Commission's air quality standards limits PM$_{10}$ (PM$<$10 $\mu$m in diameter) concentrations to 50 $\mu$g m$^{-3}$ average over 24 hours and 40 $\mu$g m$^{-3}$ over a year \citep{ECStandards15}.  

Measurements made with instruments at monitoring stations are considered authoritative; however, these observations are often too sparse to deliver regional maps at sufficient resolution to assess progress with mitigation strategies and for monitoring compliance. One solution is to couple spatially sparse monitoring station observations with spatially complete chemistry transport model (CTM) output, \citep[see, e.g.,][]{vdKassteeleS06a,DenbyEtAl08a,CandianiEtAl13a}. In such settings, monitoring station observations serve as a regression model outcome with CTM output set as a predictor. 

\begin{equation}\label{eq:pmRegression}
  \text{PM}_{10}(\bs) = \beta_0 + w_0(\bs) + \text{CTM}(\bs) \left\{\beta_{CTM} + w_{CTM}(\bs)\right\} + \epsilon(\bs)\;.
\end{equation}

This illustration draws on data and analyses presented in \citep{hamm2015,datta2016}. We consider April 6, 2010, PM$_{10}$ measurements across central Europe with corresponding output from the LOTOS-EUROS~\citep{SchaapEtAl08a} CTM. Following \citep{hamm2015} we hypothesis a space-varying relationship between the PM$_{10}$ measurements observed at monitoring stations and CTM output. In what follows, we compare fit metrics for three candidate models derived from (\ref{eq:pmRegression}): 1) a non-spatial regression; 2) space-varying intercept; 3) space-varying intercept and CTM output. Resulting model objects are called \code{pm.1}, \code{pm.2}, and \code{pm.3}, respectively. For brevity, code only for fitting \code{pm.3} is shown. We then consider parameter estimates and associated plots of the spatial random effects from \code{pm.3}, followed by development of predictive maps of both the space-varying coefficients and PM$_{10}$ prediction for a grid over the study area.

We begin by loading the data and separating it into a ``model'' set \code{PM10.mod} comprising locations where both PM$_{10}$ measurements and CTM values are available, and a ``prediction'' set \code{PM10.pred} where only CTM values are available. Here too, we calculated the maximum distance between any two monitoring stations which will help with setting prior distributions for spatial decay parameters.

\begin{knitrout}
\definecolor{shadecolor}{rgb}{0.969, 0.969, 0.969}\color{fgcolor}\begin{kframe}
\begin{alltt}
\hlkwd{data}\hlstd{(PM10.dat)}

\hlstd{PM10.mod} \hlkwb{<-} \hlstd{PM10.dat[}\hlopt{!}\hlkwd{is.na}\hlstd{(PM10.dat}\hlopt{$}\hlstd{pm10.obs),]}
\hlstd{PM10.pred} \hlkwb{<-} \hlstd{PM10.dat[}\hlkwd{is.na}\hlstd{(PM10.dat}\hlopt{$}\hlstd{pm10.obs),]}

\hlstd{d.max} \hlkwb{<-} \hlkwd{max}\hlstd{(}\hlkwd{iDist}\hlstd{(PM10.mod[,}\hlkwd{c}\hlstd{(}\hlstr{"x.coord"}\hlstd{,}\hlstr{"y.coord"}\hlstd{)]))}
\hlstd{d.max} \hlcom{#km}
\end{alltt}
\begin{verbatim}
[1] 2929.193
\end{verbatim}
\end{kframe}
\end{knitrout}

The code below specifies the model covariance parameters' prior distributions, and MCMC sampler starting and Metropolis proposal variance values. Unlike the simulated data analysis, here we demonstrate placing independent GPs on the intercept and CTM predictor. This requires priors for a process specific spatial decay parameter $\phi$ and variance $\sigma^2$. We again use a Uniform prior for the process' decay parameters that provides support for an effective spatial range between $\sim$ 3 and 2197 km, given an exponential covariance function. The two spatial variances and single observational variance $\tau^2$ each are assumed to follow an IG with shape 2 and scale 1. We center the IG's on 1, because it is approximately equal to the residual variance from the first candidate model, i.e., the non-spatial regression. One should generally do careful exploratory data analysis to arrive at a robust set of prior distributions and hyperparameters

\begin{knitrout}
\definecolor{shadecolor}{rgb}{0.969, 0.969, 0.969}\color{fgcolor}\begin{kframe}
\begin{alltt}
\hlstd{r} \hlkwb{<-} \hlnum{2}

\hlstd{priors} \hlkwb{<-} \hlkwd{list}\hlstd{(}\hlstr{"phi.Unif"}\hlstd{=}\hlkwd{list}\hlstd{(}\hlkwd{rep}\hlstd{(}\hlnum{3}\hlopt{/}\hlstd{(}\hlnum{0.75}\hlopt{*}\hlstd{d.max), r),} \hlkwd{rep}\hlstd{(}\hlnum{3}\hlopt{/}\hlstd{(}\hlnum{0.001}\hlopt{*}\hlstd{d.max), r)),}
               \hlstr{"sigma.sq.IG"}\hlstd{=}\hlkwd{list}\hlstd{(}\hlkwd{rep}\hlstd{(}\hlnum{2}\hlstd{, r),} \hlkwd{rep}\hlstd{(}\hlnum{1}\hlstd{, r)),}
               \hlstr{"tau.sq.IG"}\hlstd{=}\hlkwd{c}\hlstd{(}\hlnum{2}\hlstd{,} \hlnum{1}\hlstd{))}

\hlstd{starting} \hlkwb{<-} \hlkwd{list}\hlstd{(}\hlstr{"phi"}\hlstd{=}\hlkwd{rep}\hlstd{(}\hlnum{3}\hlopt{/}\hlstd{(}\hlnum{0.1}\hlopt{*}\hlstd{d.max), r),} \hlstr{"sigma.sq"}\hlstd{=}\hlkwd{rep}\hlstd{(}\hlnum{1}\hlstd{, r),} \hlstr{"tau.sq"}\hlstd{=}\hlnum{1}\hlstd{)}

\hlstd{tuning} \hlkwb{<-} \hlkwd{list}\hlstd{(}\hlstr{"phi"}\hlstd{=}\hlkwd{rep}\hlstd{(}\hlnum{0.1}\hlstd{, r),} \hlstr{"sigma.sq"}\hlstd{=}\hlkwd{rep}\hlstd{(}\hlnum{0.05}\hlstd{, r),} \hlstr{"tau.sq"}\hlstd{=}\hlnum{0.1}\hlstd{)}

\hlstd{n.samples} \hlkwb{<-} \hlnum{10000}

\hlstd{m.3} \hlkwb{<-} \hlkwd{spSVC}\hlstd{(pm10.obs} \hlopt{~} \hlstd{pm10.ctm,} \hlkwc{coords}\hlstd{=}\hlkwd{c}\hlstd{(}\hlstr{"x.coord"}\hlstd{,}\hlstr{"y.coord"}\hlstd{),}
             \hlkwc{data}\hlstd{=PM10.mod,} \hlkwc{starting}\hlstd{=starting,} \hlkwc{svc.cols}\hlstd{=}\hlkwd{c}\hlstd{(}\hlnum{1}\hlstd{,}\hlnum{2}\hlstd{),}
             \hlkwc{tuning}\hlstd{=tuning,} \hlkwc{priors}\hlstd{=priors,} \hlkwc{cov.model}\hlstd{=}\hlstr{"exponential"}\hlstd{,}
             \hlkwc{n.samples}\hlstd{=n.samples,} \hlkwc{n.report}\hlstd{=}\hlnum{5000}\hlstd{,} \hlkwc{n.omp.threads}\hlstd{=}\hlnum{4}\hlstd{)}
\end{alltt}
\begin{verbatim}
----------------------------------------
	General model description
----------------------------------------
Model fit with 256 observations.

Number of covariates 2.

Number of space varying covariates 2.

Using the exponential spatial correlation model.

Number of MCMC samples 10000.

Priors and hyperpriors:
	beta flat.
	Diag(K) IG hyperpriors
		parameter	shape		scale
		K[1,1]		2.000000	1.000000
		K[2,2]		2.000000	1.000000

	phi Unif lower bound hyperpriors:	0.001	0.001	
	phi Unif upper bound hyperpriors:	1.024	1.024	

	tau.sq IG hyperpriors shape=2.00000 and scale=1.00000

Source compiled with OpenMP, posterior sampling is using 4 thread(s).
-------------------------------------------------
		Sampling
-------------------------------------------------
Sampled: 5000 of 10000, 50.00%
Report interval Metrop. Acceptance rate: 36.84%
Overall Metrop. Acceptance rate: 36.84%
-------------------------------------------------
Sampled: 10000 of 10000, 100.00%
Report interval Metrop. Acceptance rate: 36.08%
Overall Metrop. Acceptance rate: 36.46%
-------------------------------------------------
\end{verbatim}
\end{kframe}
\end{knitrout}

We again pass the \code{spSVC} object to \code{spRecover} for composition sampling of the remaining model parameters needed for posterior summaries, model assessment, and subsequent prediction. 

\begin{knitrout}
\definecolor{shadecolor}{rgb}{0.969, 0.969, 0.969}\color{fgcolor}\begin{kframe}
\begin{alltt}
\hlstd{m.3} \hlkwb{<-} \hlkwd{spRecover}\hlstd{(m.3,} \hlkwc{start}\hlstd{=}\hlkwd{floor}\hlstd{(}\hlnum{0.75}\hlopt{*}\hlstd{n.samples),} \hlkwc{thin}\hlstd{=}\hlnum{2}\hlstd{,}
                 \hlkwc{n.omp.threads}\hlstd{=}\hlnum{4}\hlstd{,} \hlkwc{verbose}\hlstd{=}\hlnum{FALSE}\hlstd{)}
\end{alltt}
\end{kframe}
\end{knitrout}

Passing the \code{spRecover} object to \code{spDiag} yields several popular model fit diagnostics, two of which are summarized in Tables~\ref{tab:dic} and \ref{tab:gp}. Table~\ref{tab:dic} shows the deviance information criterion (DIC) and associated effective number of parameters pD \citep{Spiegelhalter01}, while Table~\ref{tab:gp} presents a posterior predictive loss metric D = G+P proposed by \citep{GG98}, where G measures goodness of fit and P penalizes complexity. Models with lower values of DIC or D are preferred over those with higher values. Both metrics favor Model 3 which allows both the intercept and CTM predictor to vary spatially over the study area.

\begin{minipage}{.45\textwidth}
\centering

\begin{tabular}{lrr}
\toprule
  & pD & DIC\\
\midrule
Model 1 & 2.99 & 363.35\\
Model 2 & 84.61 & 188.88\\
Model 3 & 160.53 & 81.55\\
\bottomrule
\end{tabular}

\captionof{table}{Model fit using DIC.\label{tab:dic}}
\end{minipage}
\begin{minipage}{.45\textwidth}
\centering

\begin{tabular}{lrrr}
\toprule
  & G & P & D\\
\midrule
Model 1 & 380.53 & 389.12 & 769.65\\
Model 2 & 94.23 & 189.33 & 283.56\\
Model 3 & 22.74 & 122.94 & 145.68\\
\bottomrule
\end{tabular}

\captionof{table}{Model fit using GPD.\label{tab:gp}}
\end{minipage}

Again, passing \code{spRecover}'s \code{coda} objects to \code{summary} provides posterior summaries of regression coefficients and covariance parameters. 
\begin{knitrout}
\definecolor{shadecolor}{rgb}{0.969, 0.969, 0.969}\color{fgcolor}\begin{kframe}
\begin{alltt}
\hlkwd{round}\hlstd{(}\hlkwd{summary}\hlstd{(m.3}\hlopt{$}\hlstd{p.beta.recover.samples)}\hlopt{$}\hlstd{quantiles[,}\hlkwd{c}\hlstd{(}\hlnum{3}\hlstd{,}\hlnum{1}\hlstd{,}\hlnum{5}\hlstd{)],}\hlnum{3}\hlstd{)}
\end{alltt}
\begin{verbatim}
              50%   2.5% 97.5%
(Intercept) 3.189  2.105 4.286
pm10.ctm    0.324 -0.087 0.726
\end{verbatim}
\begin{alltt}
\hlkwd{round}\hlstd{(}\hlkwd{summary}\hlstd{(m.3}\hlopt{$}\hlstd{p.theta.recover.samples)}\hlopt{$}\hlstd{quantiles[,}\hlkwd{c}\hlstd{(}\hlnum{3}\hlstd{,}\hlnum{1}\hlstd{,}\hlnum{5}\hlstd{)],}\hlnum{3}\hlstd{)}
\end{alltt}
\begin{verbatim}
                       50%  2.5% 97.5%
sigma.sq.(Intercept) 0.278 0.146 0.480
sigma.sq.pm10.ctm    0.103 0.066 0.153
tau.sq               0.286 0.137 0.463
phi.(Intercept)      0.426 0.072 0.909
phi.pm10.ctm         0.001 0.001 0.002
\end{verbatim}
\end{kframe}
\end{knitrout}

\begin{figure}[ht!]
\centering
\subfigure[$\tbeta(\bs)_0$]{\includegraphics[trim={0cm 0cm 0cm 2cm},clip,width=3in]{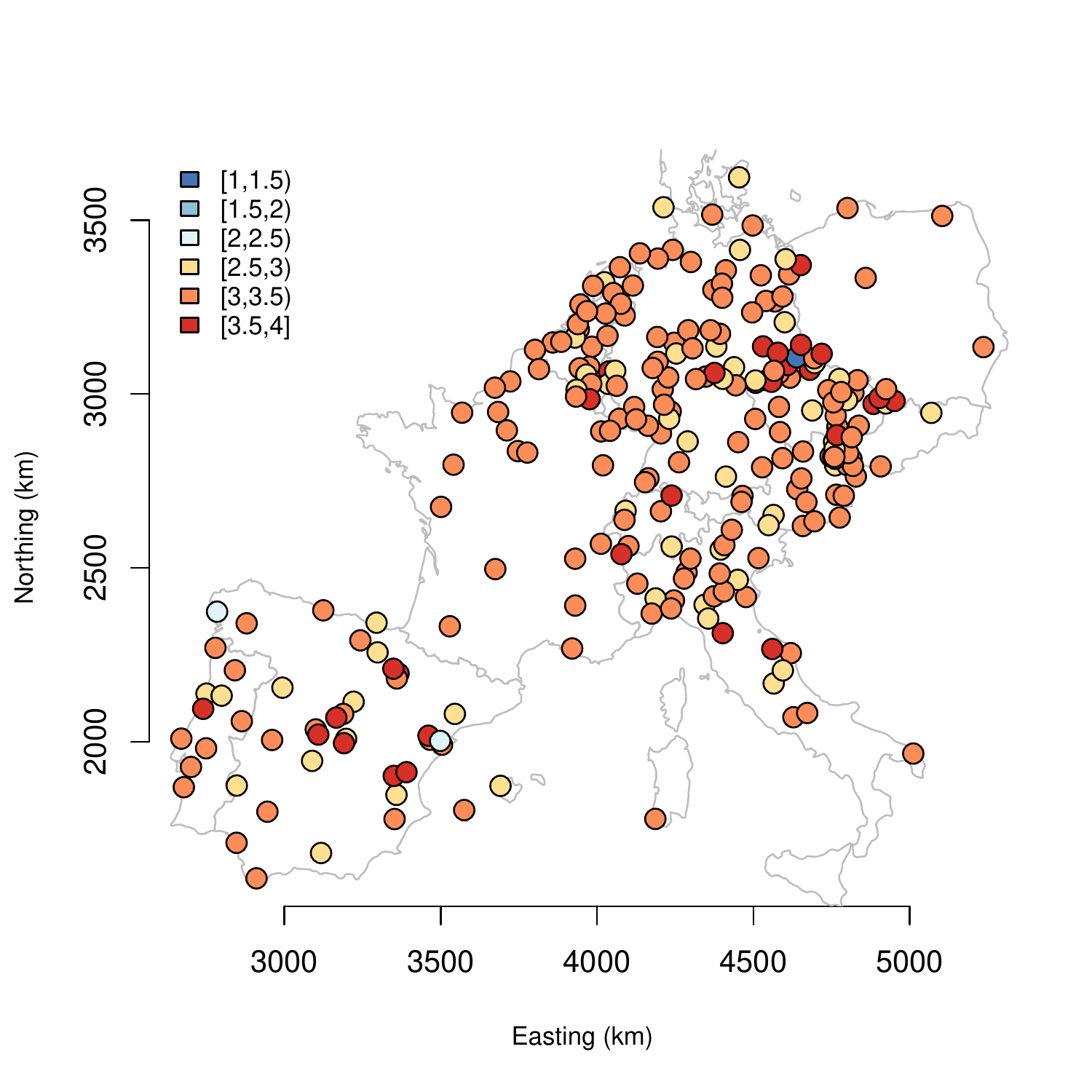}\label{pm10Beta0}} 
\subfigure[$\tbeta(\bs)_{CTM}$]{\includegraphics[trim={0cm 0cm 0cm 2cm},clip,width=3in]{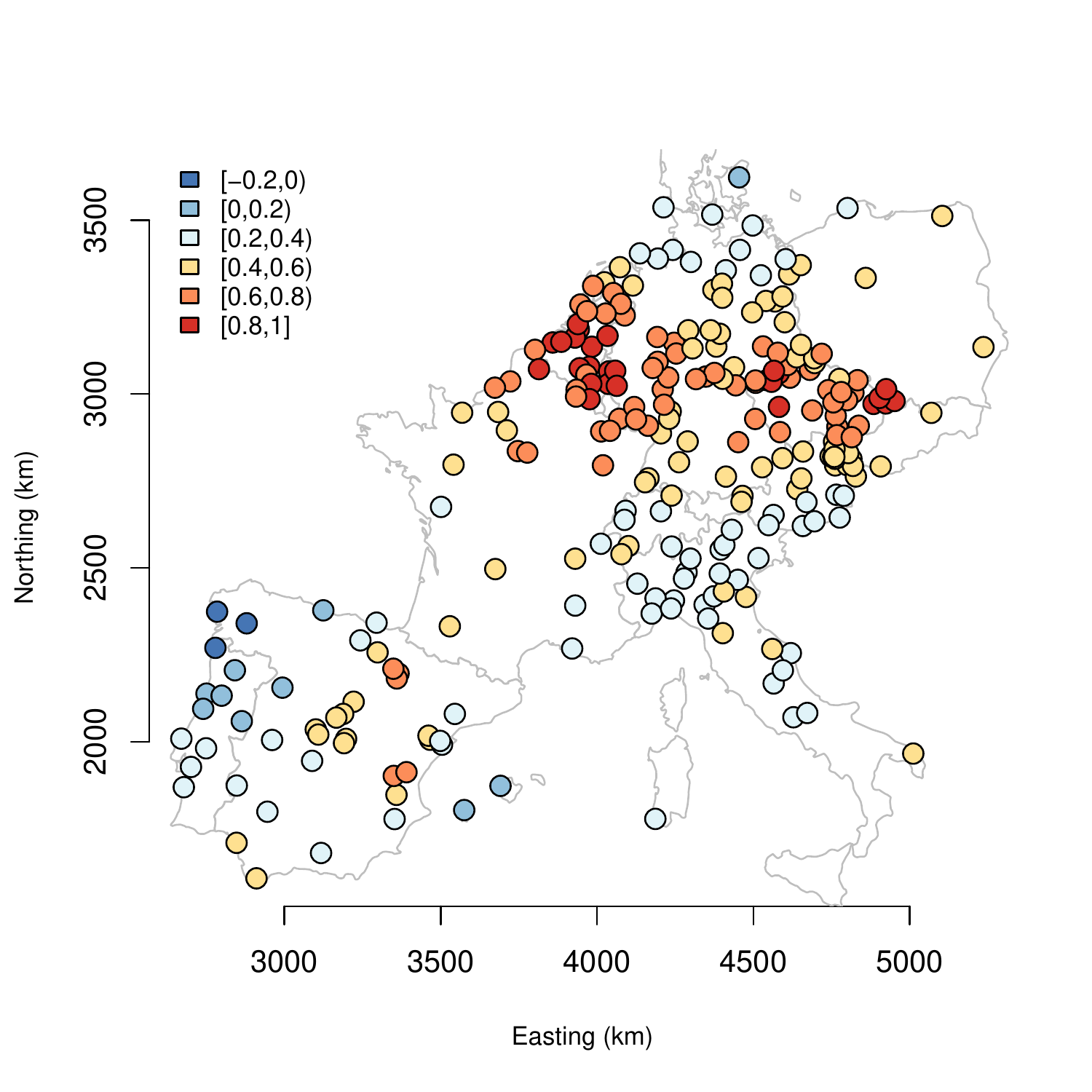}\label{pm10BetaCTM}} \\
\subfigure[Predicted PM$_{10}$]{\includegraphics[trim={0cm 0cm 0cm 2cm},clip,width=3in]{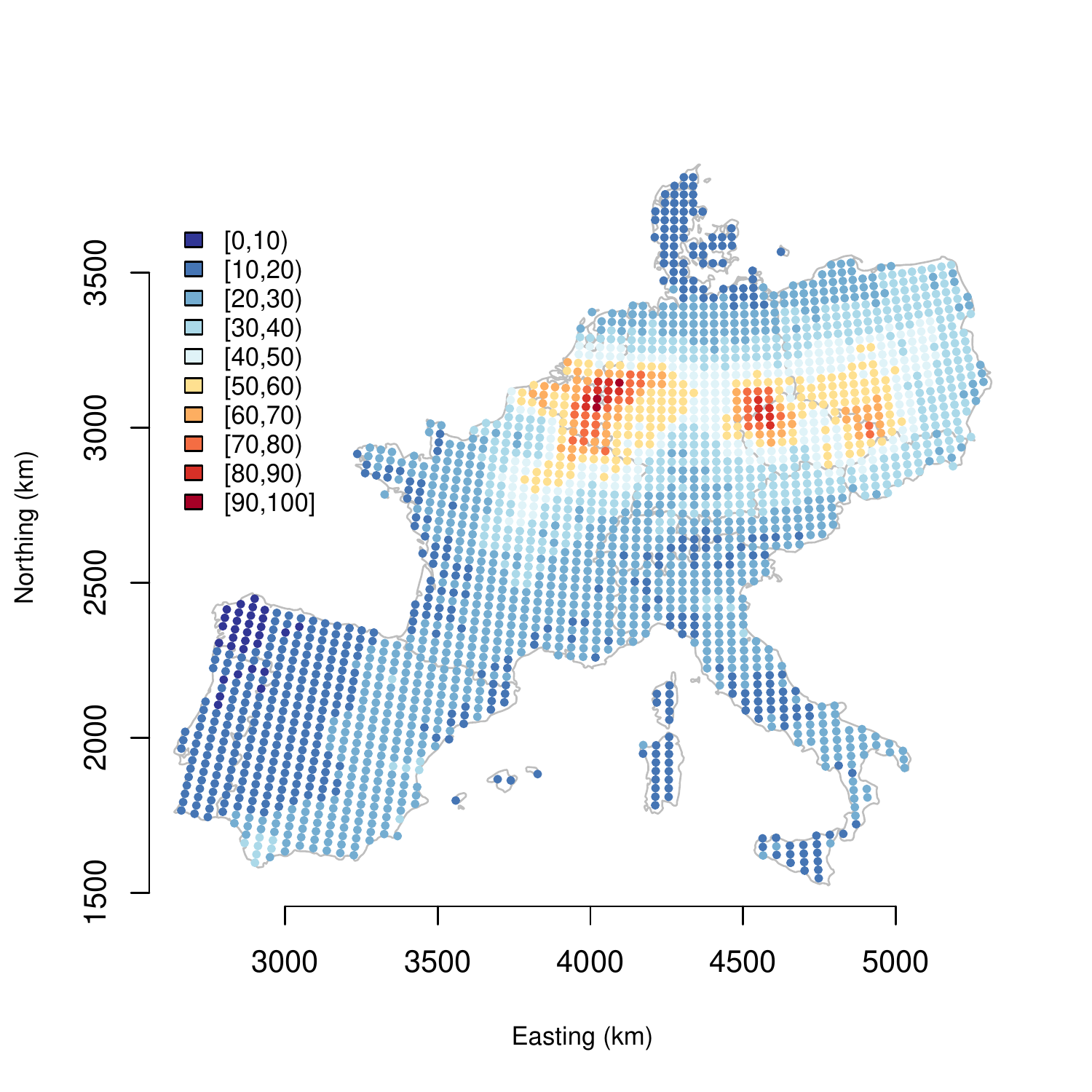}\label{pm10Pred}}
\subfigure[Probability of PM$_{10}>50$ $\mu$g m$^{-3}$]{\includegraphics[trim={0cm 0cm 0cm 2cm},clip,width=3in]{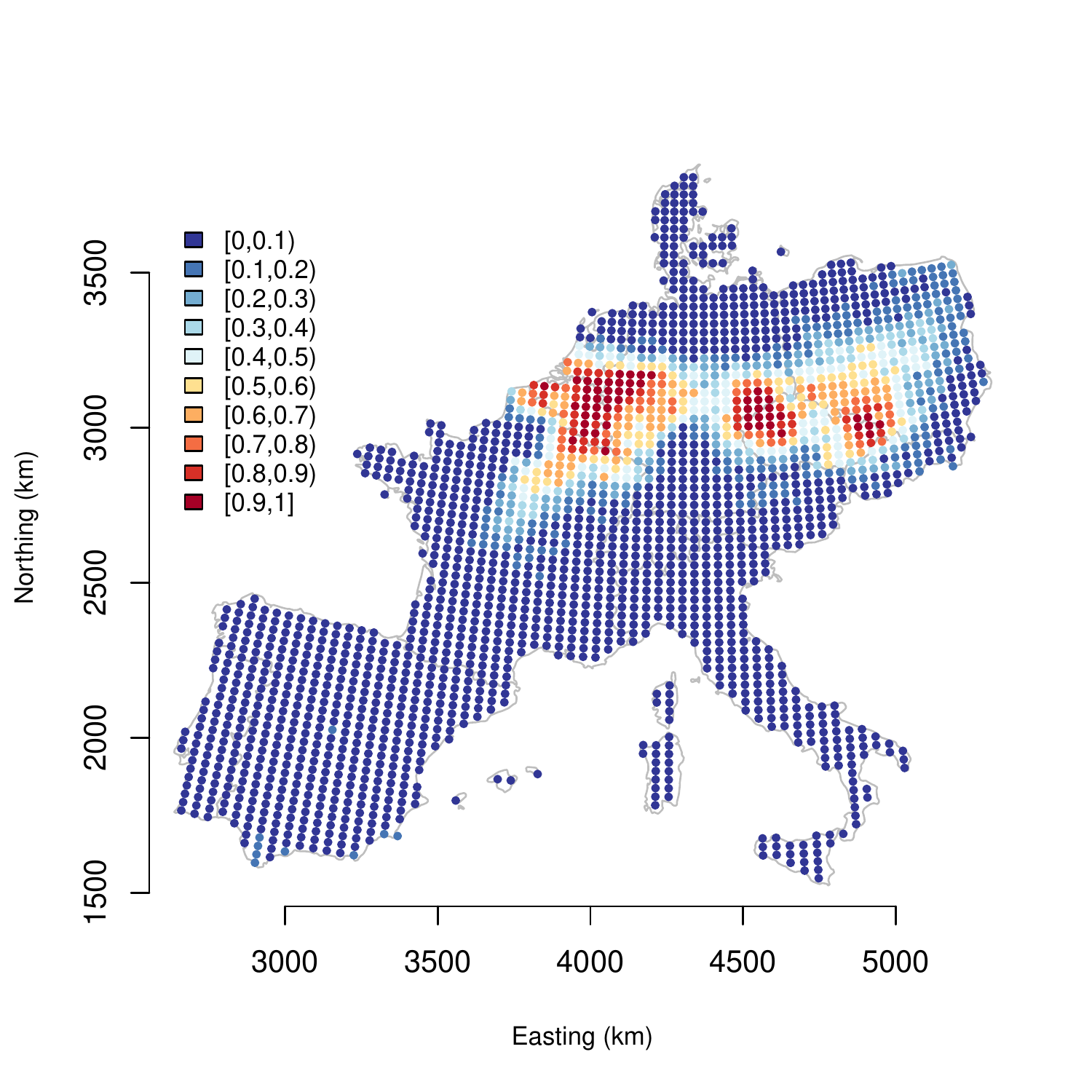}\label{pm10PredGr50}}
\caption{Estimated space-varying intercept \subref{pm10Beta0} and CTM regression coefficient \subref{pm10BetaCTM} over observed monitoring locations. PM$_{10}$ posterior predictive distribution median \subref{pm10Pred} and probability of regulatory exceedance \subref{pm10PredGr50}. }\label{pm10Maps}
\end{figure}

Given the spatial decay parameter estimates, the corresponding effective spatial range (defined as the distance at which the correlation drops to 0.05) posterior median and 95\% CI for the intercept and CTM processes are approximately 7.03 (3.3, 41.82) km and 2005.4 (1330, 2183.12) km, respectively. While the CTM predictor does have a long spatial range relative to the size of the study area, model fit metrics and the magnitude of its process variance \code{sigma.sq.pm10.ctm} estimates relative to the intercept process and nugget variance, offer evidence for a space-varying relationship with the outcome variable. This conclusion is further reinforced by Figure~\ref{pm10BetaCTM}, which shows the posterior median for the CTM predictor regression coefficient, $\tilde{\beta}_{CTM}(\bs)$'s, over observed monitoring locations. These posterior samples, along with those of the space-varying intercept, $\tilde{\beta}_{0}(\bs)$'s, are extracted from \code{m.3} and summarized in the code below (\code{tilde.beta.0} and \code{tilde.beta.ctm} are displayed in Figure~\ref{pm10Beta0}-\subref{pm10BetaCTM}).

\begin{knitrout}
\definecolor{shadecolor}{rgb}{0.969, 0.969, 0.969}\color{fgcolor}\begin{kframe}
\begin{alltt}
\hlstd{tilde.beta.0} \hlkwb{<-} \hlkwd{apply}\hlstd{(}
    \hlstd{m.3}\hlopt{$}\hlstd{p.tilde.beta.recover.samples[[}\hlstr{"tilde.beta.(Intercept)"}\hlstd{]],}
    \hlnum{1}\hlstd{, median)}

\hlstd{tilde.beta.ctm} \hlkwb{<-} \hlkwd{apply}\hlstd{(}
    \hlstd{m.3}\hlopt{$}\hlstd{p.tilde.beta.recover.samples[[}\hlstr{"tilde.beta.pm10.ctm"}\hlstd{]],}
    \hlnum{1}\hlstd{, median)}
\end{alltt}
\end{kframe}
\end{knitrout}

We next turn to prediction over the grid of 2336 CTM output locations via a call to \code{spPredict}. As illustrated below, this call uses samples from a \code{spRecover} object along with the prediction locations (\code{pred.coords}) and associated design matrix (\code{pred.covars}). The argument \code{joint} specifies if posterior predictive samples should be drawn from the joint or point-wise distribution. 

\begin{knitrout}
\definecolor{shadecolor}{rgb}{0.969, 0.969, 0.969}\color{fgcolor}\begin{kframe}
\begin{alltt}
\hlstd{m.3.pred} \hlkwb{<-} \hlkwd{spPredict}\hlstd{(m.3,} \hlkwc{pred.covars}\hlstd{=}\hlkwd{cbind}\hlstd{(}\hlnum{1}\hlstd{, PM10.pred}\hlopt{$}\hlstd{pm10.ctm),}
                      \hlkwc{pred.coords}\hlstd{=PM10.pred[,}\hlnum{1}\hlopt{:}\hlnum{2}\hlstd{],} \hlkwc{thin}\hlstd{=}\hlnum{25}\hlstd{,}
                      \hlkwc{joint}\hlstd{=}\hlnum{TRUE}\hlstd{,} \hlkwc{n.omp.threads}\hlstd{=}\hlnum{4}\hlstd{,} \hlkwc{verbose}\hlstd{=}\hlnum{FALSE}\hlstd{)}
\end{alltt}
\end{kframe}
\end{knitrout}

If the number of prediction locations is large, joint prediction can be prohibitively expensive. Even here with 2336 locations, 51 samples, and using 4 CPUs, joint posterior sampling takes 3.53 minutes verses 0.72 minutes for point-wise sampling.

Joint prediction results are given in the bottom row of Figure~\ref{pm10Maps}. The posterior predictive distribution median map (Figure~\ref{pm10Pred}) shows three distinct zones of high PM$_{10}$ values over Central Europe. A compelling quality of MCMC-based inference is access to the posterior predictive distribution. This access facilitates summaries like that given in Figure~\ref{pm10PredGr50} which identifies the probability that a given location will exceed a PM$_{10}$ value 50 $\mu$g m$^{-3}$ (as further explored in \cite{hamm2015} and \cite{datta2016}).

\section{Summary} \label{sec:summary}
The new \code{spSVC} function more fully implements the computationally efficient MCMC algorithm detailed in \cite{FBG15} and provides a flexible software tool for fitting spatially varying coefficient models. While other software, some of which are noted in Section~\ref{sec:intro}, offer similar spatially adaptive regression, few provide both univariate and multivariate GP specifications and the computational efficiency delivered by the proposed sampling algorithm and use of \proglang{OpenMP} parallelization in combination with optional calls to multi-lower-level BLAS and LAPACK multi-threaded matrix algebra libraries. Future work will focus on extending this function to accommodate non-Guassian and multivariate outcomes, as well as for settings where the number of locations precluded the use of full-rank spatial GPs.

\clearpage
\section*{Acknowledgments}
Finley was supported by National Science Foundation (NSF) EF-1253225 and DMS-1916395, and National Aeronautics and Space Administration's Carbon Monitoring System project. Banerjee was supported by NSF DMS-1513654, IIS-1562303, and DMS-1916349.


\begin{thebibliography}{44}
\newcommand{\enquote}[1]{``#1''}
\providecommand{\natexlab}[1]{#1}
\providecommand{\url}[1]{\texttt{#1}}
\providecommand{\urlprefix}{URL }
\expandafter\ifx\csname urlstyle\endcsname\relax
  \providecommand{\doi}[1]{doi:\discretionary{}{}{}#1}\else
  \providecommand{\doi}{doi:\discretionary{}{}{}\begingroup
  \urlstyle{rm}\Url}\fi
\providecommand{\eprint}[2][]{\url{#2}}

\bibitem[{Bakar \emph{et~al.}(2015{\natexlab{a}})Bakar, Kokic, and
  Jin}]{spTDyn-JSCS}
Bakar KS, Kokic P, Jin H (2015{\natexlab{a}}).
\newblock \enquote{Hierarchical spatially varying coefficient and temporal
  dynamic process models using {spTDyn}.}
\newblock \emph{Journal of Statistical Computation and Simulation}.
\newblock \urlprefix\url{10.1080/00949655.2015.1038267}.

\bibitem[{Bakar \emph{et~al.}(2015{\natexlab{b}})Bakar, Kokic, and
  Jin}]{spTDyn-JRSS}
Bakar KS, Kokic P, Jin H (2015{\natexlab{b}}).
\newblock \enquote{A spatio-dynamic model for assessing frost risk in
  south-eastern Australia.}
\newblock \emph{Journal of the Royal Statistical Society, Series C}.
\newblock \urlprefix\url{10.1111/rssc.12103}.

\bibitem[{Bakar \emph{et~al.}(2017)Bakar, Kokic, and Jin}]{spTDyn}
Bakar KS, Kokic P, Jin H (2017).
\newblock \emph{Spatially varying and spatio-temporal dynamic linear models}.
\newblock R package version 2.0.

\bibitem[{Bakar and Sahu(2018)}]{spTimer}
Bakar KS, Sahu SK (2018).
\newblock \emph{Spatio-Temporal Bayesian Modeling}.
\newblock R package version 3.3.

\bibitem[{Bakka \emph{et~al.}(2018)Bakka, Rue, Fuglstad, Riebler, Bolin,
  Illian, Krainski, Simpson, and Lindgren}]{Bakka19}
Bakka H, Rue H, Fuglstad GA, Riebler AI, Bolin D, Illian J, Krainski E, Simpson
  DP, Lindgren FK (2018).
\newblock \enquote{Spatial modelling with INLA: A review.}
\newblock \emph{ArXiv e-prints}.
\newblock \eprint{1802.06350}.

\bibitem[{Banerjee \emph{et~al.}(2014)Banerjee, Carlin, and Gelfand}]{BCG14}
Banerjee S, Carlin BP, Gelfand AE (2014).
\newblock \emph{Hierarchical Modeling and Analysis for Spatial Data}.
\newblock Second edition. Chapman \& Hall/CRC, Boca Raton, FL.

\bibitem[{Bivand(2019)}]{CRANSP}
Bivand R (2019).
\newblock \emph{CRAN Task View: Analysis of Spatial Data}.
\newblock 2019-02-25,
  \urlprefix\url{https://cran.r-project.org/web/views/Spatial.html}.

\bibitem[{Bivand and Yu(2017)}]{spgwr}
Bivand R, Yu D (2017).
\newblock \emph{spgwr: Geographically Weighted Regression}.
\newblock R package version 0.6-32,
  \urlprefix\url{https://CRAN.R-project.org/package=spgwr}.

\bibitem[{Brunekreef and Holgate(2002)}]{BrunekreefH02a}
Brunekreef B, Holgate ST (2002).
\newblock \enquote{Air Pollution and Health.}
\newblock \emph{The Lancet}, \textbf{360}(9341), 1233--1242.

\bibitem[{B\"{u}rgin and Ritschard(2017)}]{vcrpart}
B\"{u}rgin R, Ritschard G (2017).
\newblock \enquote{Coefficient-Wise Tree-Based Varying Coefficient Regression
  with vcrpart.}
\newblock \emph{Journal of Statistical Software, Articles}, \textbf{80}(6),
  1--33.
\newblock ISSN 1548-7660.
\newblock \doi{10.18637/jss.v080.i06}.

\bibitem[{Candiani \emph{et~al.}(2013)Candiani, Carnevale, Finzi, Pisoni, and
  Volta}]{CandianiEtAl13a}
Candiani G, Carnevale C, Finzi G, Pisoni E, Volta M (2013).
\newblock \enquote{A Comparison of Reanalysis Techniques: Applying Optimal
  Interpolation and Ensemble {K}alman Filtering to Improve Air Quality
  Monitoring at Mesoscale.}
\newblock \emph{Science of the Total Environment}, \textbf{458-460}(0), 7--14.

\bibitem[{Carpenter \emph{et~al.}(2017)Carpenter, Gelman, Hoffman, Lee,
  Goodrich, Betancourt, Brubaker, Guo, Li, and Riddell}]{stan}
Carpenter B, Gelman A, Hoffman M, Lee D, Goodrich B, Betancourt M, Brubaker M,
  Guo J, Li P, Riddell A (2017).
\newblock \enquote{Stan: A Probabilistic Programming Language.}
\newblock \emph{Journal of Statistical Software, Articles}, \textbf{76}(1),
  1--32.
\newblock ISSN 1548-7660.
\newblock \doi{10.18637/jss.v076.i01}.
\newblock \urlprefix\url{https://www.jstatsoft.org/v076/i01}.

\bibitem[{Dagum and Menon(1998)}]{openmp98}
Dagum L, Menon R (1998).
\newblock \enquote{OpenMP: an industry standard API for shared-memory
  programming.}
\newblock \emph{Computational Science \& Engineering, IEEE}, \textbf{5}(1),
  46--55.

\bibitem[{Datta \emph{et~al.}(2016)Datta, Banerjee, Finley, Hamm, and
  Schaap}]{datta2016}
Datta A, Banerjee S, Finley A, Hamm N, Schaap M (2016).
\newblock \enquote{Nonseparable dynamic nearest neighbor Gaussian process
  models for large spatio-temporal data with an application to particulate
  matter analysis.}
\newblock \emph{Annals of Applied Statistics}, \textbf{10}(3), 1286--1316.
\newblock ISSN 1932-6157.

\bibitem[{Denby \emph{et~al.}(2008)Denby, Schaap, Segers, Builtjes, and
  Horalek}]{DenbyEtAl08a}
Denby B, Schaap M, Segers A, Builtjes P, Horalek J (2008).
\newblock \enquote{Comparison of Two Data Assimilation Methods for Assessing
  {PM10} Exceedances on the {E}uropean Scale.}
\newblock \emph{Atmospheric Environment}, \textbf{42}(30), 7122--7134.

\bibitem[{{European Commission}(2015)}]{ECStandards15}
{European Commission} (2015).
\newblock \enquote{European Union Air Quality Standards.}
\newblock \emph{http://ec.europa.eu/environment/air/quality/standards.htm}.

\bibitem[{Fan and Zhang(2008)}]{Fan08}
Fan J, Zhang W (2008).
\newblock \enquote{Statistical Methods with Varying Coefficient Models.}
\newblock \emph{Statistics and its interface}, \textbf{1 1}, 179--195.

\bibitem[{Finley \emph{et~al.}(2015)Finley, Banerjee, and Gelfand}]{FBG15}
Finley A, Banerjee S, Gelfand A (2015).
\newblock \enquote{spBayes for Large Univariate and Multivariate
  Point-Referenced Spatio-Temporal Data Models.}
\newblock \emph{Journal of Statistical Software, Articles}, \textbf{63}(13),
  1--28.
\newblock ISSN 1548-7660.

\bibitem[{Fotheringham \emph{et~al.}(2002)Fotheringham, Brunsdon, and
  Charlton}]{Fotheringham02}
Fotheringham A, Brunsdon C, Charlton M (2002).
\newblock \emph{Geographically Weighted Regression: The Analysis of Spatially
  Varying Relationships}.
\newblock Wiley.
\newblock ISBN 9780471496168.

\bibitem[{Gelfand and Ghosh(1998)}]{GG98}
Gelfand AE, Ghosh SK (1998).
\newblock \enquote{{Model choice: A minimum posterior predictive loss
  approach}.}
\newblock \emph{Biometrika}, \textbf{85}(1), 1--11.

\bibitem[{Gelfand \emph{et~al.}(2003{\natexlab{a}})Gelfand, Kim, Sirmans, and
  Banerjee}]{Gelfand03}
Gelfand AE, Kim HJ, Sirmans CF, Banerjee S (2003{\natexlab{a}}).
\newblock \enquote{Spatial Modeling With Spatially Varying Coefficient
  Processes.}
\newblock \emph{Journal of the American Statistical Association},
  \textbf{98}(462), 387--396.

\bibitem[{Gelfand \emph{et~al.}(2003{\natexlab{b}})Gelfand, Kim, Sirmans, and
  Banerjee}]{GKSB03}
Gelfand AE, Kim HJ, Sirmans CF, Banerjee S (2003{\natexlab{b}}).
\newblock \enquote{Spatial Modeling With Spatially Varying Coefficient
  Processes.}
\newblock \emph{Journal of the American Statistical Association},
  \textbf{98}(462), 387--396.

\bibitem[{Gelman \emph{et~al.}(2013)Gelman, Carlin, Stern, Dunson, Vehtari, and
  Rubin}]{gelman2013}
Gelman A, Carlin J, Stern H, Dunson D, Vehtari A, Rubin D (2013).
\newblock \emph{Bayesian Data Analysis, Third Edition}.
\newblock Chapman \& Hall/CRC Texts in Statistical Science. Taylor \& Francis.
\newblock ISBN 9781439840955.

\bibitem[{Hamm \emph{et~al.}(2015)Hamm, Finley, Schaap, and Stein}]{hamm2015}
Hamm N, Finley A, Schaap M, Stein A (2015).
\newblock \enquote{A Spatially Varying Coefficient Model for Mapping PM10 Air
  Quality at the European scale.}
\newblock \emph{Atmospheric Environment}, \textbf{102}, 393--405.

\bibitem[{Hastie and Tibshirani(1993)}]{Hastie93}
Hastie T, Tibshirani R (1993).
\newblock \enquote{Varying-Coefficient Models.}
\newblock \emph{Journal of the Royal Statistical Society. Series B
  (Methodological)}, \textbf{55}(4), 757--796.

\bibitem[{Hayfield and Racine(2008)}]{np}
Hayfield T, Racine JS (2008).
\newblock \enquote{Nonparametric Econometrics: The np Package.}
\newblock \emph{Journal of Statistical Software}, \textbf{27}(5).
\newblock \urlprefix\url{http://www.jstatsoft.org/v27/i05/}.

\bibitem[{Heim(2007)}]{svcm}
Heim S (2007).
\newblock \emph{svcm: 2d and 3d space-varying coefficient models in R}.
\newblock R package version 0.1.2.

\bibitem[{Helske(2019)}]{walker}
Helske J (2019).
\newblock \emph{{walker}: Bayesian Regression with Time-Varying Coefficients}.
\newblock R package version 0.2.4-1,
  \urlprefix\url{http://github.com/helske/walker}.

\bibitem[{Henderson and Searle(1981)}]{henderson1981}
Henderson HV, Searle SR (1981).
\newblock \enquote{On deriving the inverse of a sum of matrices.}
\newblock \emph{SIAM Review}, \textbf{23}(1), 53--60.

\bibitem[{Hoek \emph{et~al.}(2013)Hoek, Krishnan, Beelen, Peters, Ostro,
  Brunekreef, and Kaufman}]{HoekEtAl13a}
Hoek G, Krishnan RM, Beelen R, Peters A, Ostro B, Brunekreef B, Kaufman JD
  (2013).
\newblock \enquote{Long-term air pollution exposure and cardio- respiratory
  mortality: a review.}
\newblock \emph{Environmental Health}, \textbf{12}, 43.

\bibitem[{Hothorn \emph{et~al.}(2018)Hothorn, Buehlmann, Kneib, Schmid, and
  Hofner}]{mboost}
Hothorn T, Buehlmann P, Kneib T, Schmid M, Hofner B (2018).
\newblock \emph{{mboost}: Model-Based Boosting}.
\newblock {R} package version 2.9-1,
  \urlprefix\url{https://CRAN.R-project.org/package=mboost}.

\bibitem[{Lindgren and Rue(2015)}]{Lindgren15}
Lindgren F, Rue H (2015).
\newblock \enquote{Bayesian Spatial Modelling with R-INLA.}
\newblock \emph{Journal of Statistical Software}, \textbf{63}(19), 1--25.
\newblock \urlprefix\url{http://www.jstatsoft.org/v63/i19/}.

\bibitem[{Loomis \emph{et~al.}(2013)Loomis, Grosse, Lauby-Secretan,
  El~Ghissassi, Bouvard, Benbrahim-Tallaa, Guha, Baan, Mattock, and
  Straif}]{LoomisEtAl13a}
Loomis D, Grosse Y, Lauby-Secretan B, El~Ghissassi F, Bouvard V,
  Benbrahim-Tallaa L, Guha N, Baan R, Mattock H, Straif S (2013).
\newblock \enquote{The Carcinogenicity of Outdoor Air Pollution.}
\newblock \emph{The Lancet Oncology}, \textbf{14}(13), 1262--1263.

\bibitem[{{R Core Team}(2018)}]{R}
{R Core Team} (2018).
\newblock \emph{R: A Language and Environment for Statistical Computing}.
\newblock R Foundation for Statistical Computing, Vienna, Austria.
\newblock \urlprefix\url{https://www.R-project.org/}.

\bibitem[{Robert and Casella(2004)}]{robert2010}
Robert C, Casella G (2004).
\newblock \emph{Monte Carlo Statistical Methods}.
\newblock Springer Texts in Statistics, second edition. Springer-Verlag.

\bibitem[{Roberts and Rosenthal(2009)}]{Roberts09}
Roberts GO, Rosenthal JS (2009).
\newblock \enquote{Examples of Adaptive MCMC.}
\newblock \emph{Journal of Computational and Graphical Statistics},
  \textbf{18}(2), 349--367.

\bibitem[{Rue \emph{et~al.}(2009)Rue, Martino, and Chopin}]{INLA}
Rue H, Martino S, Chopin N (2009).
\newblock \enquote{Approximate {Bayesian} Inference for Latent {Gaussian}
  Models Using Integrated Nested {Laplace} Approximations (with discussion).}
\newblock \emph{Journal of the Royal Statistical Society B}, \textbf{71},
  319--392.

\bibitem[{Schaap \emph{et~al.}(2008)Schaap, Timmermans, Roemer, Boersen,
  Builtjes, Sauter, Velders, and Beck}]{SchaapEtAl08a}
Schaap M, Timmermans RMA, Roemer M, Boersen GAC, Builtjes P, Sauter F, Velders
  G, Beck J (2008).
\newblock \enquote{The {LOTOS-EUROS} Model: Description, Validation and Latest
  Developments.}
\newblock \emph{International Journal of Environment and Pollution},
  \textbf{32}(2), 270--290.

\bibitem[{Spiegelhalter \emph{et~al.}(2001)Spiegelhalter, Best, Carlin, and
  van~der Linde}]{Spiegelhalter01}
Spiegelhalter DJ, Best NG, Carlin BP, van~der Linde A (2001).
\newblock \enquote{Bayesian Measures of Model Complexity and Fit.}

\bibitem[{{Stan Development Team}(2018)}]{RStan}
{Stan Development Team} (2018).
\newblock \enquote{{RStan}: the {R} interface to {Stan}.}
\newblock R package version 2.18.2, \urlprefix\url{http://mc-stan.org/}.

\bibitem[{van~de Kassteele and Stein(2006)}]{vdKassteeleS06a}
van~de Kassteele J, Stein A (2006).
\newblock \enquote{A Model for External Drift Kriging with Uncertain Covariates
  applied to Air Quality Measurements and Dispersion Model Output.}
\newblock \emph{Environmetrics}, \textbf{17}(4), 309--322.

\bibitem[{Vihola \emph{et~al.}(2017)Vihola, Helske, and Franks}]{walker-arxiv}
Vihola M, Helske J, Franks J (2017).
\newblock \enquote{Importance Sampling Type Estimators Based on Approximate
  Marginal MCMC.}
\newblock \emph{ArXiv e-prints}.
\newblock \eprint{1609.02541}.

\bibitem[{Wood(2017)}]{wood04}
Wood S (2017).
\newblock \emph{Generalized Additive Models: An Introduction with R}.
\newblock 2 edition. Chapman and Hall/CRC.

\bibitem[{Zhang(2016)}]{zhang13}
Zhang X (2016).
\newblock \enquote{An Optimized BLAS Library Based on GotoBLAS2.}
\newblock \url{https://github.com/xianyi/OpenBLAS/}.
\newblock Accessed 2015-06-01.

\end{thebibliography}
\end{document}